\documentclass{emulateapj}
\usepackage{apjfonts}
\usepackage{natbib}          
\shorttitle{OPTICAL PROPERTIES OF THE LMXB-GC CONNECTION}
\shortauthors{SIVAKOFF ET AL.}

\defcitealias{JCF+2004}{ACSVCS3}
\defcitealias{BIS+2006}{B06}

\newcommand\gdiff[1]{{#1}}

\accepted{by {\it Astrophysical Journal}, January 20, 2007}
\slugcomment{}

\begin{document}

\title{The Low-Mass X-ray Binary and Globular Cluster Connection in Virgo Cluster Early-type Galaxies: Optical Properties}

\author{
Gregory R. Sivakoff\altaffilmark{1,2},
Andr\'{e}s Jord\'{a}n\altaffilmark{3},
Craig L. Sarazin\altaffilmark{1},
John P. Blakeslee\altaffilmark{4},
Patrick C{\^ o}t{\' e}\altaffilmark{5},
Laura Ferrarese\altaffilmark{5},
Adrienne M. Juett\altaffilmark{1},
Simona Mei\altaffilmark{6,7}, and
Eric W. Peng\altaffilmark{5}
}

\altaffiltext{1}{
Department of Astronomy,
University of Virginia,
P. O. Box 400325,
Charlottesville, VA 22904-4325, USA;
grs8g@virginia.edu,
sarazin@virginia.edu,
ajuett@virginia.edu}
\altaffiltext{2}{
Current Address:
Department of Astronomy,
The Ohio State University,
4055 McPherson Laboratory
140 W. 18th Avenue, Columbus, OH 43210-1173, USA;
sivakoff@astronomy.ohio-state.edu}
\altaffiltext{3}{
European Southern Observatory,
Karl-Schwarzschild-Str.\ 2
85748 Garching bei M\"{u}nchen, Germany;
ajordan@eso.org}
\altaffiltext{4}{
Department of Physics and Astronomy,
Washington State University, 
1245 Webster Hall, Pullman, WA 99163-2814; 
jblakes@wsu.edu}
\altaffiltext{5}{
Herzberg Institute of Astrophysics,
5071 W. Saanic Road,
Victoria, BC V9E 2E7, Canada;
patrick.cote@nrc-cnrc.gc.ca,
laura.ferrarese@nrc-cnrc.gc.ca,
eric.peng@nrc-cnrc.gc.ca}
\altaffiltext{6}{
GEPI,
Observatoire de Paris,
Section de Meudon,
5 Place  J.Janss  en,
92195 Meudon Cedex,
France;
Simona.Mei@obspm.fr
}
\altaffiltext{7}{
Department of Physics and Astronomy,
Johns Hopkins University,
Baltimore, MD 21218, USA
}

\begin{abstract}
Low-mass X-ray binaries (LMXBs) form efficiently in globular clusters
(GCs). By combining {\it Chandra X-ray Observatory} and {\it Hubble
Space Telescope}-Advanced Camera for Surveys observations of
early-type galaxies, we probe the LMXB-GC connection using the most
accurate identification of LMXBs and GCs to date. We explore the
optical properties of 270 GCs with LMXBs and 6,488 GCs without
detectable X-ray emission from a sample of eleven massive early-type
galaxies in the Virgo cluster. Globular clusters that are more
massive, are redder, and have {\em smaller radii} are more likely to
contain LMXBs. Using known structural scaling relations for GCs, the
latter implies that denser GCs are more likely to hold LMXBs. Unlike
Galactic GCs, a large number of GCs with LMXBs have half-mass
relaxation times $>2.5 {\rm \, Gyr}$; GCs do not need to survive for
more than five relaxation timescales to produce LMXBs. By fitting the
dependence of the expected number of LMXBs per GC, $\lambda_{t}$, on
the GC mass $\gdiff{M}$, color $(g-z)$, and half-mass radius $r_{h,
{\rm cor}}$, we find that $\lambda_{t} \propto \gdiff{M}^{1.24\pm0.08}
\, 10^{0.9^{+0.2}_{-0.1} \, (g-z)} \, r_{h, {\rm
cor}}^{-2.2\pm0.2}$. This rules out that the number of LMXBs
per GC is linearly proportional to GC mass (99.89\% confidence
limit). We derive an expression to estimate the number of multiple
LMXB sources in GCs and predict that most GCs with high X-ray
luminosities contain a single LMXB. The detailed dependence of
$\lambda_{t}$ on GC properties appears mainly due to a dependence on a
combination of mass and radius, and a dependence on color, that are
essentially equivalent to a dependence on the encounter rate
$\Gamma_h$ and the metallicity $\gdiff{Z}$, $ \lambda_{t} \propto
\Gamma_{h}^{0.82\pm0.05} \, \gdiff{Z}^{0.39\pm0.07} $. Our analysis
provides strong evidence that dynamical formation and metallicity play
the primary roles in determining the presence of an LMXB in
extragalactic GCs.
The shallower than linear dependence
of GC sources requires an explanation by theories of dynamical binary
formation; however, we note that our use of $\Gamma_{h}$ as a proxy
for the encounter rate, particularly if core-collapsed extragalactic
GCs preferentially contain LMXBs, needs further testing in nearer
galaxies.
A metallicity-dependent variation in the number of neutron stars and
black holes per unit GC mass, effects from irradiation induced
winds, or suppression of magnetic braking in metal-poor stars
may all be consistent with our derived abundance dependence;
all three scenarios require further development.
\end{abstract}
\keywords{
binaries: close ---
galaxies: elliptical and lenticular, cD ---
galaxies: star clusters ---
globular clusters: general ---
X-rays: binaries ---
X-rays: galaxies
}

\section{Introduction}

A low-mass X-ray binary (LMXB) consists of a compact stellar remnant,
a neutron star (NS) or black hole (BH), that accretes the stellar
envelope of a low-mass ($\gdiff{M} \lesssim 2 \gdiff{M}_\odot$)
stellar companion. Roche-lobe overflow is the major method of mass
transfer in these systems, which in the Milky Way have orbital periods
of $0.19$--$398 {\rm \, hr}$ \citep{WNP1995}. When actively accreting,
LMXBs are bright X-ray sources ($\sim 10^{35}$--$10^{39} {\rm \, erg
\, s}^{-1}$; hereafter, the use of ``LMXBs'' refers only to active
LMXBs). The Milky Way ($L_* \sim 10^{10} \, L_\odot$) contains
$\approx 150$ active LMXBs \citep{LPH2001}. Most of these LMXBs are in
the field of the Milky Way and are likely to have been the result of
the stellar evolution of a primordial binary \citep{VH1995}. This
primordial formation of an LMXB cannot be particularly efficient as
the binary must survive the supernova that forms the NS/BH, yet remain
in a tight enough orbit that Roche-lobe overflow eventually occurs.

On the other hand, gravitational interactions in dense stellar
systems, particularly globular clusters (GCs), could lead to more
efficient formation of binaries in general, and LMXBs in
particular. Over the history of X-ray astronomy, 13 active Galactic
LMXBs have been detected in 12 GCs \citep{LPH2001,WA2001}. In
addition to active LMXBs, Galactic GCs may contain seven times as many
LMXBs in quiescence \citep{HGL+2003}. Since GCs account for $\sim
0.1\%$ of the Galactic light, but $\sim 10\%$ of the active Galactic
LMXBs, formation of LMXBs in GCs must be hundreds of times more
efficient than in the field \citep[e.g.,][]{C1975,K1975}. The
increased efficiency may be due to a combination of two effects: (1)
LMXB formation through tidal capture or exchange interactions after
the supernova of the NS/BH; and (2) tighter binary orbits (binary
hardening) in GCs due to exchange interactions. The optical properties
of GCs containing LMXBs provide a platform to test the dynamical
formation of LMXBs in GCs. Furthermore, the metallicities of GCs can
be readily estimated through their optical color, allowing one to
probe for metallicity dependence in the evolution or formation
processes of LMXBs in GCs. Although some patterns can be seen in the
limited Galactic data \citep[e.g.,][hereafter
\citetalias{BIS+2006}]{BPF+1995,BIS+2006}, a larger sample is
necessary to probe LMXB formation and evolution in GCs.

With the launch of the {\it Chandra X-ray Observatory} ({\it CXO}), it
is now possible to examine large samples of LMXBs. In particular,
X-ray observations of nearby early-type galaxies can identify tens to
hundreds of bright X-ray sources, most of which are likely to be LMXBs
\citep[e.g.,][]{SIB2000}. In these galaxies, $\sim 20\%$--$70\%$ of
the LMXBs appear to be associated with GCs
\citep[e.g.,][]{SIB2000,SIB2001,ALM2001,KMZ2002,JCF+2004}; early-type
galaxies can provide the large number of GCs containing LMXBs
necessary to probe LMXB formation and evolution in GCs. Early samples
have shown that brighter and redder GCs are more likely to contain
LMXBs \citep[e.g.][]{KMZ+2003,SKI+2003,JCF+2004}; however, the use of
resolution limited ground-based data or field-of-view (FOV) limited
space-based data suppressed the numbers of GCs that could clearly be
identified.

By observing nearby early-type galaxies with the Wide Field Channel of
the {\it Hubble Space Telescope} Advanced Camera for Surveys ({\it
HST}-ACS/WFC), which has a $3\farcm2 \times 3\farcm2$ FOV and can
resolve GCs in nearby galaxies, one can construct accurate and
comprehensive lists of GCs. In particular, the measurement of GC size
allows better discrimination against unresolved background sources
with GC colors, while providing a GC property that is crucially
important for probing the structure of GCs that contain LMXBs. The
ACS Virgo Cluster Survey (ACSVCS) has observed the centers of 100
early-type galaxies \citep{CBF+2004} and provides excellent GC
data. Some of its initial work explored the GC-LMXB connection in
NGC~4486 \citetext{M87, \citealp{JCF+2004}, hereafter
\citetalias{JCF+2004}}. In this work, we expand upon this analysis
exploring the GC-LMXB connection in 10 ACSVCS galaxies and a more
nearby galaxy (NGC~4697) that may not belong to Virgo, but was
observed following the ACSVCS observational setup. We identify 270 GCs
that contain LMXBs; this is one of the largest such samples to date,
and the largest sample that includes information about GC size.

\begin{deluxetable}{lccccc}
\tabletypesize{\footnotesize}
\tablewidth{-232.41493pt}
\tablecaption{Galaxy Properties of Sample\label{tab:acsvcs_gclmxb_gal_sample}}
\tablehead{
&
Type &
$r_{\rm eff}$ &
$D_{\rm SBF}$ &
$M_B$&
\\
Galaxy&
(RC3) &
(arcsec) &
(Mpc) &
(mag) &
Other Names\\
(1)&
(2)&
(3)&
(4)& 
(5)}
\startdata
\object{NGC 4365} & E3 & \phn49.8 & 23.2 & -21.46 & VCC731\\
\object{NGC 4374} & E1 & \phn51.0 & 18.7 & -21.45 & M84, VCC763\\
\object{NGC 4382} & S0 & \phn54.6 & 17.9 & -21.46 & M85, VCC798\\
\object{NGC 4406} & E3 &    104.0 & 18.0 & -21.56 & M86, VCC881\\
\object{NGC 4472} & E2 &    104.0 & 16.8 & -21.98 & M49, VCC1226\\
\object{NGC 4486} & E0 & \phn94.9 & 15.7 & -21.70 & M87, VCC1316\\
\object{NGC 4526} & S0 & \phn44.4 & 16.1 & -20.60 & VCC1535\\
\object{NGC 4552} & E0 & \phn29.3 & 16.1 & -20.55 & M89, VCC1632\\
\object{NGC 4621} & E5 & \phn40.5 & 15.0 & -20.37 & M59, VCC1903\\
\object{NGC 4649} & E2 & \phn68.7 & 16.4 & -21.39 & M60, VCC1978\\
\object{NGC 4697} & E6 & \phn72.0 & 11.3 & -20.29 &
\enddata
\end{deluxetable}

\section{Sample}

\subsection{Galaxy Sample}

We selected our sample from early-type Virgo galaxies with detections
of LMXBs using {\it CXO} and GCs using {\it HST}-ACS. Since the most
massive galaxies are the most likely to have large populations of both
LMXBs and GCs, we concentrated on galaxies with $M_B < -20.25$. Our
sample includes the ten brightest ACSVCS galaxies (NGC 4472, 4486,
4406, 4365, 4382, 4374, 4649, 4526, 4552, and 4621) and NGC~4697, a
similarly bright galaxy, which we observed with nearly the same setup
as the ACSVCS. Table~\ref{tab:acsvcs_gclmxb_gal_sample} summarizes
some properties of the galaxies in our sample. The first three columns
list the name, Hubble type, and effective radius $r_{\rm eff}$ from
\citet{VVC+1992}. The next columns list the (polynomial calibrated)
surface brightness fluctuation distance, $D_{\rm SBF}$, determined
from the {\it HST} observations \citep{MBC+2006} and the absolute
total B-magnitude, $M_B$ (combining $b_t$ from HyperLeda\footnote{See
\url{http://leda.univ-lyon1.fr/}.} and $D_{\rm SBF}$).

\subsection{X-ray Analysis}

\begin{deluxetable}{lcccc}
\tablewidth{-178.35136pt}
\tablecaption{Properties of {\itshape Chandra} Observations \label{tab:acsvcs_gclmxb_xobs_sample}}
\tablehead{
&
&
&
$t_{\rm obs}$ & 
$t_{\rm exp}$\\
Galaxy&
OBSID &
Detector\tablenotemark{a} &
(ks)&
(ks)\\
(1)&
(2)&
(3)&
(4)& 
(5)}
\startdata
NGC 4365  & \dataset[ADS/Sa.CXO#obs/02015]{2015} & ACIS-S3/F  &   40.4 &   40.4\\
NGC 4374  & \dataset[ADS/Sa.CXO#obs/00803]{0803} & ACIS-S3/VF &   28.5 &   28.4\\
NGC 4382  & \dataset[ADS/Sa.CXO#obs/02016]{2016} & ACIS-S3/F  &   39.7 &   39.7\\
NGC 4406  & \dataset[ADS/Sa.CXO#obs/00318]{0318} & ACIS-S3/F  &   14.6 &   11.4\\ 
\nodata   & \dataset[ADS/Sa.CXO#obs/00963]{0963} & ACIS-S3/F  &   14.8 &   12.0\\
NGC 4472  & \dataset[ADS/Sa.CXO#obs/00321]{0321} & ACIS-S3/VF &   39.6 &   34.5\\
\nodata   & \dataset[ADS/Sa.CXO#obs/00322]{0322} & ACIS-I/VF  &   10.4 &\phn7.2\\
NGC 4486  & \dataset[ADS/Sa.CXO#obs/00352]{0352} & ACIS-S3/G  &   37.7 &   33.6\\
\nodata   & \dataset[ADS/Sa.CXO#obs/02707]{2707} & ACIS-S3/F  &   98.7 &   82.2\\
\nodata   & \dataset[ADS/Sa.CXO#obs/03717]{3717} & ACIS-S3/F  &   20.6 &\phn9.2\\
NGC 4526  & \dataset[ADS/Sa.CXO#obs/03925]{3925} & ACIS-S3/VF &   43.5 &   41.5\\
NGC 4552  & \dataset[ADS/Sa.CXO#obs/02072]{2072} & ACIS-S3/VF &   54.4 &   54.4\\
NGC 4621  & \dataset[ADS/Sa.CXO#obs/02068]{2068} & ACIS-S3/F  &   24.8 &   24.8\\
NGC 4649  & \dataset[ADS/Sa.CXO#obs/00785]{0785} & ACIS-S3/VF &   36.9 &   17.0\\
NGC 4697  & \dataset[ADS/Sa.CXO#obs/00784]{0784} & ACIS-S3/F  &   39.3 &   37.2\\
\nodata   & \dataset[ADS/Sa.CXO#obs/04727]{4727} & ACIS-S3/VF &   39.9 &   39.9\\
\nodata   & \dataset[ADS/Sa.CXO#obs/04727]{4727} & ACIS-S3/VF &   35.6 &   35.6\\
\nodata   & \dataset[ADS/Sa.CXO#obs/04729]{4729} & ACIS-S3/VF &   38.1 &   32.0\\
\nodata   & \dataset[ADS/Sa.CXO#obs/04730]{4730} & ACIS-S3/VF &   40.0 &   40.0
\enddata
\end{deluxetable}
\tablenotetext{a}{The data modes, faint mode (F), graded mode (G) and very faint mode (VF) follow the detector used.}

These galaxies all had archival or proprietary data from {\it CXO};
however, the observational setup varied widely between
galaxies. Table~\ref{tab:acsvcs_gclmxb_xobs_sample} summarizes the
properties of the {\it CXO} observations, listing the galaxy name,
{\it CXO} observation number (OBSID), detector with data mode (Faint,
Graded, or Very-Faint), and the exposure times before ($t_{\rm obs}$)
and after removing periods of high backgrounds ($t_{\rm exp}$). Since
the ACIS-I chips are less sensitive than the ACIS-S3 chip, we only
include data from ACIS-I when its $t_{\rm obs}$ is $>25\%$ the $t_{\rm
obs}$ of the ACIS-S3. Since the {\it HST-ACS} FOV is smaller than the
{\it CXO} FOV, we hereafter only discuss X-ray sources within the {\it
CXO/HST} overlapping region.

All {\it CXO} observations were analyzed under {\sc ciao 3.1}%
\footnote{See \url{http://asc.harvard.edu/ciao/}.}
with {\sc caldb 2.28} and NASA's {\sc FTOOLS 5.3}%
\footnote{See
\url{http://heasarc.gsfc.nasa.gov/docs/software/lheasoft/}%
\label{ftn:acs_vcs_o_heasoft}.}. For Observation 0784, the focal plane
temperature was $-110^\circ$ C, while all other observations were
taken at $-120^\circ$ C. When Very-Faint mode telemetry was available,
the observations were cleaned using the extra data available in this
mode to reduce the background level. All other observations were
reduced in Faint mode, including OBSID 0352, which was telemetered in
``Graded'' mode. Known aspect offsets were applied for each
observation. Our analysis includes only events with ASCA grades of 0,
2, 3, 4, and 6. Photon energies were determined using the gain files
acisD2000-01-29gain\_ctiN0001.fits, except for OBSIDs 0352
(acisD2000-07-06gainN0003.fits) and 0784
(acisD1999-09-16gainN0005.fits). When appropriate, we corrected for
time dependence of the gain and the charge-transfer inefficiency. All
observations were corrected for quantum efficiency (QE) degradation
and had exposure maps determined at $750 {\rm \, eV}$. We excluded bad
pixels, bad columns, and columns adjacent to bad columns or chip node
boundaries.

The use of local backgrounds in point source analysis mitigates the
effect of high background periods (``background flares''), which
especially affect the backside-illuminated S1 and S3
chips%
\footnote{See
\url{http://cxc.harvard.edu/contrib/maxim/acisbg/}%
\label{ftn:acs_vcs_o_bkg}.}. We avoided the periods with the most
extreme flares by only including times where the background rate was
below three times the expected rate from the blank-sky background
fields in the CALDB. When available, we used the S1 chip to measure
the background rate, excluding point sources, and compared to the
blank-sky background count rates in CALDB in the $2.6$--$6.0 {\rm \,
keV}$ band. If the S1 chip was not available, we used the S3 chip to
determine count rates in the $2.6$--$7.0 {\rm \, keV}$ band. For
Observation 0784, we compared count rates in the 0.3--10.0 keV band to
Maxim Markevitch's aciss\_B\_7\_bg\_evt\_271103.fits blank-sky
background\footnote{See footnote \ref{ftn:acs_vcs_o_bkg}.}. When we
binned the observations regularly in time to determine the count rate,
small time intervals at the edges of the existing good-time intervals
could not be included in a bin. This led to only a small amount of
time that was lost in any observation. We list the exposure before
and after flare removal in Table~\ref{tab:acsvcs_gclmxb_xobs_sample}.

To identify the discrete X-ray source population, we applied a wavelet
detection algorithm (the CIAO {\it wavdetect} program) with scales
increasing by a factor of $\sqrt{2}$ from 1 to 32 pixels. We adopted a
source detection threshold of $10^{-6}$ ($\lesssim1$ false source per
chip) for both the ACIS S3 and I chips. Source detection excluded
regions with an exposure of less than 10\% of that of the observation
(e.g., regions at the edges of chips). To maximize the signal-to-noise
(S/N), we analyzed the wavelet detection results from combined
observations when available. We detected 708 X-ray sources that were
also in the FOV of the {\it HST}-ACS observations. There are regions
of complex gas emission at the center of NGC~4486. Although detected
sources in these region may be LMXBs, they may also be compact regions
of enhanced gas emission. To avoid mistaken identification of point
sources, \citetalias{JCF+2004} excluded all X-ray sources in these
regions. We follow this procedure, excluding 33 X-ray sources.

We used the coordinate list generated by {\sc wavdetect} in ACIS
Extract 3.34%
\footnote{See
\url{http://www.astro.psu.edu/xray/docs/TARA/ae\_users\_guide.html}.}
to refine the source positions and determine source extraction
regions. This was accomplished by determining the mean positions of
events in source extraction regions consistent with the X-ray point
spread functions (PSFs) at the positions of the sources. For most
sources the median photon energy was $0.6$--$2.6 {\rm \, keV}$ and we
required the extraction region to encircle 90\% of the X-ray PSF at
$1.5 {\rm \, keV}$. We determined the PSF at either $0.3 {\rm \, keV}$
or $4.5 {\rm \, keV}$ for the few sources whose median photon energy
was either softer or harder. In a few sources where the 90\% PSF
extraction regions overlapped, we used a lower percentage of the PSF
to define the extraction region.

When there were multiple observations of a galaxy, we matched the
positions of detected sources in individual observations to those
present in the observation with the largest exposure. Absolute
astrometric corrections were applied for all sources in all galaxies
using a two-dimensional cross-correlation technique to match sources
from the Tycho-2 Catalog \citep{HFM+2000}, 2MASS Point Source and
Extended Source Catalogs\footnote{When a source appeared in both 2MASS
catalogs, the Point Source Catalog positions were used.}, and the
USNO-B Catalog \citep{MLC+2003}. An astrometric correction (separately
in RA and Dec.) was applied if the offset was statistically
significant.

We used a local background from an annular region whose area was
approximately three times that of each source's extraction region;
these local backgrounds include the diffuse emission from the host
galaxy. To insure that the local background regions did not contain
any significant number of source counts, the inner radius of the
background region was taken to be the radius encircling 97\% of the
PSF. In cases where background regions overlapped or fell along
node/chip boundaries, we slightly altered these overlapping regions,
preserving the ratio of source to background areas and ensuring that
the source region and background region had similar mean exposures.
For each of the sources, the observed net count rates, their errors,
and S/N were calculated in the $0.3$--$6.0 {\rm \, keV}$ band by
stacking the observations for each galaxy, correcting for background
photons, and dividing by the sum of the mean exposure over each source
region. In NGC 4374 and NGC 4486, a few of the sources detected with
{\sc wavdetect} had negative net count rates when determined in this
way (two sources in NGC 4374, five sources in NGC 4486); we excluded
these sources from further discussion.

\begin{deluxetable*}{lcccccccccc}
\tablewidth{-465.54189pt}
\tablecaption{GC-LMXB Matches by Galaxies \label{tab:acsvcs_gclmxb_comb_sample}}
\tablehead{
Galaxy&
$N_{X}$&
$L_{X, {\rm min}}$\tablenotemark{a}&
$N_{\rm GCs}$&
$(g-z)_{\rm div}$&
$N_{\rm Blue\,GCs}$&
$N_{\rm Red\,GCs}$&
$N_{\rm GCs\,w\,LMXBs}$&
$N_{\rm Blue\,GCs\,w\,LMXBs}$&
$N_{\rm Red\,GCs\,w\,LMXBs}$&
$N_{\rm False\,Matches}$\\
(1)&
(2)&
(3)&
(4)& 
(5)&
(6)&
(7)&
(8)& 
(9)&
(10)&
(11)}
\startdata
\multicolumn{11}{c}{Detected: All Sources With Positive Luminosity}\\
NGC 4365  &    85 & 0.10 & \phn906  & 1.160 & 418 [3] & 488 [7]\phn &    43 (5) &    11 (0) &    29 (2) & 4.36\\
NGC 4374  &    49 & 0.05 & \phn506  & 1.200 & 310 [0] & 196 [0]\phn &    18 (0) & \phn9 (0) & \phn9 (0) & 1.85\\
NGC 4382  &    37 & 0.52 & \phn505  & 1.230 & 320 [1] & 185 [1]\phn &    11 (1) & \phn7 (0) & \phn3 (0) & 1.40\\
NGC 4406  &    28 & 0.09 & \phn367  & 1.220 & 260 [0] & 107 [0]\phn & \phn6 (0) & \phn4 (0) & \phn2 (0) & 0.75\\
NGC 4472  &    86 & 0.57 & \phn764  & 1.180 & 306 [0] & 458 [4]\phn &    40 (2) & \phn4 (0) &    36 (2) & 2.83\\
NGC 4486  &    84 & 0.14 &    1639  & 1.190 & 690 [4] & 949 [10]    &    49 (6) &    12 (0) &    34 (3) & 6.95\\
NGC 4526  &    34 & 0.31 & \phn244  & 1.170 & 121 [0] & 123 [0]\phn & \phn7 (0) & \phn3 (0) & \phn4 (0) & 0.98\\
NGC 4552  &    78 & 0.15 & \phn455  & 1.190 & 217 [1] & 238 [1]\phn &    31 (1) & \phn8 (0) &    22 (0) & 3.00\\
NGC 4621  &    44 & 0.39 & \phn306  & 1.110 & 121 [0] & 185 [2]\phn &    17 (1) & \phn1 (0) &    16 (1) & 1.24\\
NGC 4649  &    60 & 0.65 & \phn806  & 1.200 & 335 [0] & 471 [0]\phn &    32 (0) & \phn3 (0) &    29 (0) & 1.50\\
NGC 4697  &    83 & 0.06 & \phn298  & 1.127 & 127 [2] & 171 [2]\phn &    34 (2) & \phn7 (1) &    27 (1) & 2.15\\
\\
\multicolumn{11}{c}{SNR: All Sources with Luminosities Determined at the 3$\sigma$ Level}\\
NGC 4365  &    29 & 2.26 & \phn906  & 1.160 & 418 [1] & 488 [5]\phn &    17 (3) & \phn1 (0) &    15 (2) & 1.25\\
NGC 4374  &    16 & 2.18 & \phn506  & 1.200 & 310 [0] & 196 [0]\phn & \phn5 (0) & \phn2 (0) & \phn3 (0) & 0.62\\
NGC 4382  &    21 & 1.64 & \phn505  & 1.230 & 320 [0] & 185 [0]\phn & \phn5 (0) & \phn4 (0) & \phn1 (0) & 0.86\\
NGC 4406  &    19 & 2.65 & \phn367  & 1.220 & 260 [0] & 107 [0]\phn & \phn6 (0) & \phn4 (0) & \phn2 (0) & 0.44\\
NGC 4472  &    54 & 1.30 & \phn764  & 1.180 & 306 [0] & 458 [2]\phn &    26 (1) & \phn3 (0) &    23 (1) & 1.72\\
NGC 4486  &    61 & 1.30 &    1639  & 1.190 & 690 [3] & 949 [5]\phn &    40 (3) & \phn9 (0) &    29 (1) & 3.65\\
NGC 4526  &    12 & 1.30 & \phn244  & 1.170 & 121 [0] & 123 [0]\phn & \phn3 (0) & \phn1 (0) & \phn2 (0) & 0.33\\
NGC 4552  &    49 & 0.80 & \phn456  & 1.190 & 217 [1] & 239 [1]\phn &    25 (1) & \phn6 (0) &    18 (0) & 1.53\\
NGC 4621  &    16 & 1.58 & \phn306  & 1.110 & 121 [0] & 185 [0]\phn & \phn4 (0) & \phn0 (0) & \phn4 (0) & 0.55\\
NGC 4649  &    21 & 2.85 & \phn807  & 1.200 & 335 [0] & 472 [0]\phn &    13 (0) & \phn2 (0) &    11 (0) & 0.43\\
NGC 4697  &    66 & 0.15 & \phn298  & 1.127 & 127 [2] & 171 [2]\phn &    26 (2) & \phn7 (1) &    19 (1) & 1.75\\
\multicolumn{11}{c}{Complete: All Sources with Luminosities $>3.2 \times 10^{38} {\rm \, erg \, s}^{-1}$}\\
NGC 4365  &    19 & 3.20 & \phn907  & 1.160 & 418 [1] & 489 [5]\phn &    10 (3) & \phn1 (0) & \phn8 (2) & 0.93\\
NGC 4374  &    12 & 3.20 & \phn506  & 1.200 & 310 [0] & 196 [0]\phn & \phn4 (0) & \phn2 (0) & \phn2 (0) & 0.45\\
NGC 4382  & \phn9 & 3.20 & \phn505  & 1.230 & 320 [0] & 185 [0]\phn & \phn1 (0) & \phn1 (0) & \phn0 (0) & 0.43\\
NGC 4406  &    16 & 3.20 & \phn367  & 1.220 & 260 [0] & 107 [0]\phn & \phn5 (0) & \phn3 (0) & \phn2 (0) & 0.38\\
NGC 4472  &    19 & 3.20 & \phn764  & 1.180 & 306 [0] & 458 [2]\phn & \phn9 (1) & \phn0 (0) & \phn9 (1) & 0.62\\
NGC 4486  &    27 & 3.20 &    1639  & 1.190 & 690 [3] & 949 [5]\phn &    14 (3) & \phn2 (0) &    10 (1) & 2.26\\
NGC 4526  & \phn4 & 3.20 & \phn244  & 1.170 & 121 [0] & 123 [0]\phn & \phn1 (0) & \phn1 (0) & \phn0 (0) & 0.11\\
NGC 4552  &    17 & 3.20 & \phn456  & 1.190 & 217 [1] & 239 [1]\phn &    10 (1) & \phn3 (0) & \phn6 (0) & 0.45\\
NGC 4621  & \phn7 & 3.20 & \phn308  & 1.110 & 121 [0] & 187 [0]\phn & \phn2 (0) & \phn0 (0) & \phn2 (0) & 0.23\\
NGC 4649  &    20 & 3.20 & \phn807  & 1.200 & 335 [0] & 472 [0]\phn &    10 (0) & \phn2 (0) & \phn8 (0) & 0.54\\
NGC 4697  & \phn3 & 3.20 & \phn298  & 1.127 & 127 [0] & 171 [0]\phn & \phn3 (0) & \phn0 (0) & \phn3 (0) & 0.00\\
\enddata
\tablenotetext{a}{Units are $10^{38} {\rm \, erg \, s}^{-1}$ in 0.3--10 keV band.}
\tablenotetext{b}{If an X-ray source was within 1\arcsec of GCs and non-GC sources, it is unclear if
the X-ray source is matched to a GC; the possibly matching GCs have
been removed from this sample.  Numbers in brackets indicate the
number of GCs where (only) multiple GCs are within 1\arcsec of an
X-ray source.}
\tablenotetext{c}{The number of RGC-LMXBs and BGC-LMXBs will not add
up to the number of GC-LMXBs if an X-ray source is within 1\arcsec of
both an RGC and a BGC. Numbers in parenthesis indicate the number of
X-ray sources where (only) multiple GCs are within 1\arcsec of an
X-ray source.}
\end{deluxetable*}

In order to convert count rates into energy fluxes, we fit a single
emission model to the spectra of all of the LMXBs in all of the
galaxies. For each observation of each galaxy, we extracted the
cumulative spectra of essentially all of the LMXBs. We did not fit
the spectra of LMXBs whose luminosities were determined at the
$\le$3$\sigma$ level or X-ray point sources that could be associated
with both a GC and a non-GC optical source. To avoid contamination
from background AGNs and foreground stars, we excluded all detected
X-ray sources that were located within $1\arcsec$ of a non-GC optical
source in the {\it HST} images. The source and background regions of
candidate LMXBs were used to extract the spectra. We binned the
spectra, requiring at least 25 total counts per bin, and only
considered bins completely in the 0.5--10.0 keV band. We
simultaneously fit the spectra of all of the sources in all of the
observations of all of the galaxies to a single emission model, which
was taken to be either a power-law or thermal bremsstrahlung. However,
we accounted for the differing Galactic absorbing columns ($N_H$) to
the different galaxies, using the Tuebingen-Boulder absorption
({\scshape tbabs}) model assuming abundances from \citet{WAM2000} and
photoelectric absorption cross-sections from \citet{VFK+1996}. Note
that the response files for each separate observation and source
include the varying effects of absorption by the contaminant that
produces the QE degradation in the ACIS detectors. The best-fit
emission model for all of the sources was found to be a bremsstrahlung
model with $kT = 9.08 ^{+0.86}_{-0.74} {\rm \, keV}$. This model,
combined with the individual values of $N_H$ and distance for each
galaxy and the individual response files for each observation, were
used to convert count rates into unabsorbed X-ray luminosities $L_X$
in the $0.3$--$10.0 {\rm \, keV}$ band.

Due mainly to varying exposure times and distances, the observations
of different galaxies have differing limiting sensitivities. For each
galaxy, we summarize the number of X-ray detections and the minimum
X-ray luminosity for three sample definitions in columns 2 and 3 of
Table~\ref{tab:acsvcs_gclmxb_comb_sample}. The three samples are: all
sources with a positive luminosity (Detected sample); all sources with
luminosities determined at the $\ge$3$\sigma$ level
(Signal-to-Noise-Ratio [SNR] sample); and all sources with $L_X \ge
3.2 \times 10^{38} {\rm \, erg \, s}^{-1}$ (Complete sample). 
All figures in this paper display the Detected sample.
Note
that the SNR sample does not include sources that were brighter than
the reported minimum luminosity but not 3$\sigma$ significant. Since a
source with 20 net counts would be detected at high completeness for
most of the galaxies, we defined the Complete sample using the
luminosity that a source in NGC~4649 (the galaxy with the highest
minimum luminosity) with 20 net counts would have, $L_X = 3.2 \times
10^{38} {\rm \, erg \, s}^{-1}$. We note that detections at the
centers of the X-ray brightest galaxies (e.g., NGC~4649) may still be
incomplete due to the presence of bright gaseous emission; however, we
have chosen this luminosity as a compromise between completeness and a
reasonable luminosity lower limit.

\subsection{Optical Analysis}

All of the galaxies in our sample, except NGC~4697, were observed as
part of the ACSVCS, which acquired two $360 {\rm \, s}$ exposures in
the F475W band ($g$-band), two $560 {\rm \, s}$ exposures in the
F850LP band, and one $90 {\rm \, s}$ F850LP exposure ($z$-band). Data
reductions were carried out as described in \citet{JBP+2004}.
NGC~4697 was observed separately from the ACSVCS, but in a similar
manner; its two F475W exposures were $15 {\rm \, s}$ longer. We
excluded optical sources detected in the central few arcseconds of
some galaxies with dusty cores (NGC 4374 and NGC 4526 in
\citealt{FCJ+2006}, and NGC 4697) and in the region of NGC~4649 that
overlaps with its nearby spiral neighbor, NGC~4647.

\subsubsection{Observed GC Properties}

\begin{figure*}
\center
\includegraphics[width=0.75\textwidth]{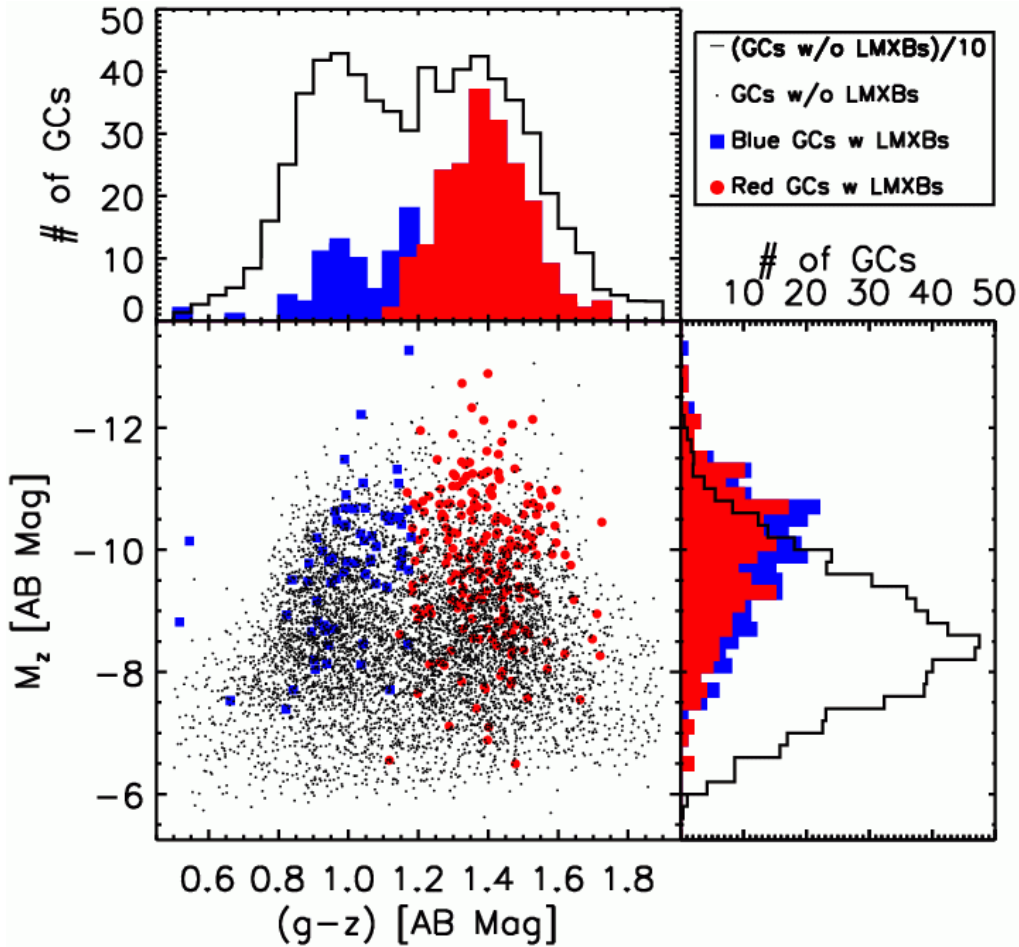}
\caption{GC magnitudes ($M_z$) versus GC colors
 ($g-z$), with integrated histograms of the properties above
and to the right. GCs unmatched to LMXBs are
indicated by small black dots and unfilled black histograms (scaled
down by a factor of 10). Blue GCs with LMXBs are indicated by filled
blue squares and histograms.  Red GCs with LMXBs are indicated by
filled red circles and histograms. The histograms of the GCs with
LMXBs are stacked on each other. GCs that are redder and brighter are
more likely to contain LMXBs.
\label{fig:acsvcs_gclmxb_cmd}}
\end{figure*}

In our sample, over 10,000 optical sources were characterized by their
magnitudes, $g$ and $z$, half-light radii, $r_{h}$, and positions as
determined by KINGPHOT \citep[see the Appendix in][where this code is described
and simulations are used to illustrate its performance]{JCB+2005}.
We note that $\sim80\%$ of the GCs in our sample have $r_{h}$ measured at ${\rm
SNR}>5$.
All magnitudes were converted
to absolute magnitudes ($M_g$ and $M_z$) using the surface brightness
fluctuation distances in Table~\ref{tab:acsvcs_gclmxb_gal_sample}.
As described in \citet{JBP+2004}, we first use magnitude, color and size
criteria to select an initial set of GC candidates. Then we use a statistical
clustering method, described in detail in another paper in the ACSVCS series
(Jord\'an et~al. 2007, in preparation), which assigns to each source in the
field of view of each galaxy a probability $p_{\rm GC}$ that the source is a GC. 
Our samples of GC candidates are then constructed by selecting all sources that
have $p_{\rm GC} \ge 0.5$. The results of our classification method are
illustrated in Figure~1 of \citet{PJC+2006}. This selection returns 7,084 likely
GCs in our sample galaxies. We list the number of GC candidates detected in each
galaxy (excluding GCs whose matching X-ray source may also match to a non-GC
optical source and all optical sources in the X-ray excluded regions of
NGC~4486) in the fourth column of Table~\ref{tab:acsvcs_gclmxb_comb_sample}.

Early-type galaxies often have a bimodal distribution of GC
colors. For each galaxy, we used the division point, $(g-z)_{\rm
div}$, in the $(g - z)$ color following \citet{PJC+2006} to divide the
GCs into blue-GCs and red-GCs; these division point ranged from
$(g-z)_{\rm div} = 1.11$ to 1.23. We list the division points and the
number of blue-GCs and red-GCs in columns five, six, and seven of
Table~\ref{tab:acsvcs_gclmxb_comb_sample}.

\begin{figure*}
\center
\includegraphics[width=0.75\textwidth]{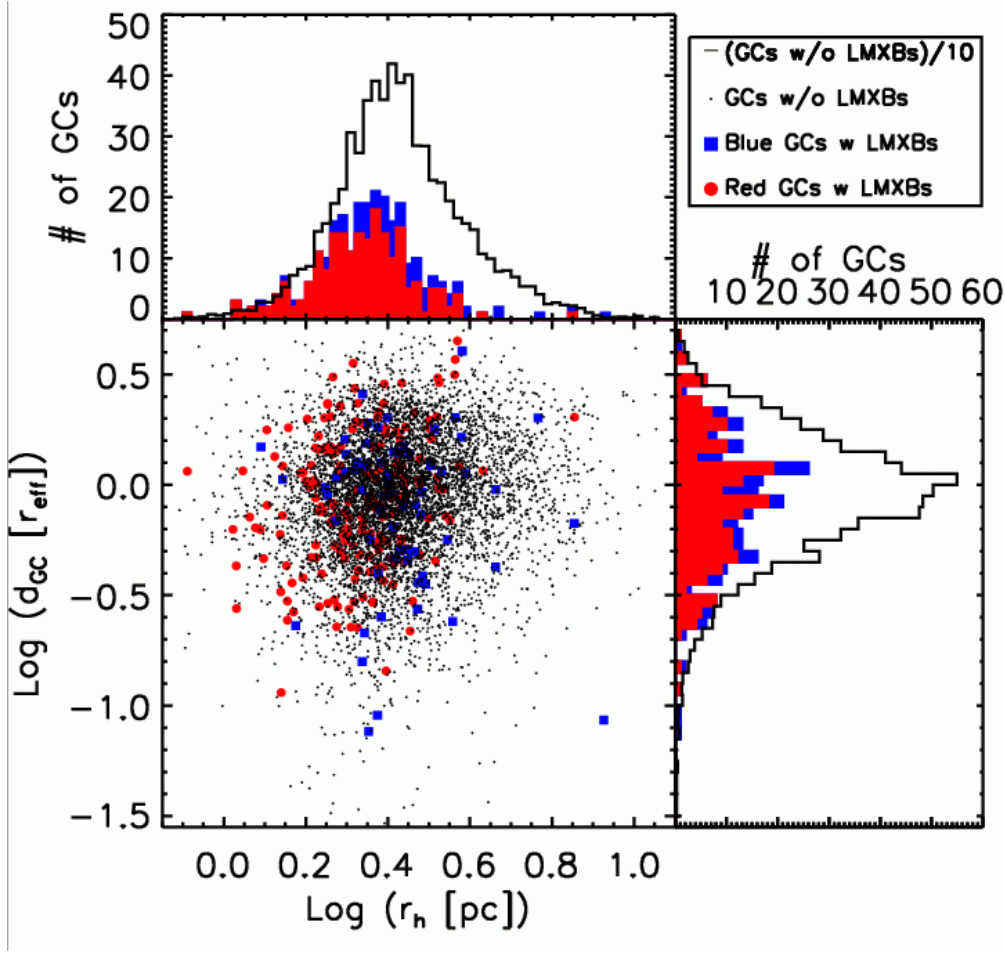}
\caption{GC galactocentric distances ($d_{\rm GC}$)
versus observed GC half-light radii ($r_{h}$), with integrated
histograms of the properties above and to the right.  The
symbols follow those of Figure~\ref{fig:acsvcs_gclmxb_cmd}.  GCs that
are smaller in extent are more likely to contain LMXBs.  The
galactocentric distance does not appear to affect whether a GC
contains an LMXB. \label{fig:acsvcs_gclmxb_size_dist}}
\end{figure*}

The positions of the GCs were used to determine the galactocentric
distance, $d_{\rm GC}$. These values were then scaled to the effective
radii of the galaxies. We show scatter plots and integrated histograms
of the observed properties of the GCs in our sample in Figures
\ref{fig:acsvcs_gclmxb_cmd} and \ref{fig:acsvcs_gclmxb_size_dist}.

\subsubsection{Derived GC Properties}

From the magnitude, colors, and radii, we can derive other parameters
that might impact the formation and evolution of LMXBs in GCs. We
calculate the GC mass ($\gdiff{M}$) directly from the $z$-band
magnitude,
\begin{equation}
  \gdiff{M} = \Upsilon_z \,
                10^{ -0.4 \, ( M_z - M_{z,\odot} ) } \,
                \gdiff{M}_\odot \,
  ,
\end{equation}
where $\Upsilon_z$ is the $z$-band mass-to-light ratio and
$M_{z,\odot} = 4.512$ is the absolute $z$-band magnitude of the Sun
obtained from {\it calcphot} in the STSDAS Synphot IRAF package. To
estimate $\Upsilon_z$, we use version 2.0 of the PEGASE code
\citep{FR1997}. In PEGASE, we used the stellar initial mass function
of \citet{K1983} to compute $(g-z)$ colors and $z$-band luminosities
for an assumed age $\tau=13$ Gyr for several fixed values of
[Fe/H]. Under these assumptions, $\Upsilon_z$ is never more than 10\%
different from 1.5 for [Fe/H] in the range $-2 \la $ [Fe/H] $ \la 0$,
and thus $z$-band magnitudes are very good tracers of the
mass. Comparably small ranges of $\Upsilon_z$ result if younger GC
ages are assumed. Given the average $(g-z)$ of the GCs in any of our
galaxies, we interpolate on these PEGASE model curves to estimate
average $z$-band mass-to-light ratios. The values of $\Upsilon_z$
depend slightly on the mean color of the GC systems, but the variation
was less than 1\% across our sample galaxies when assuming that GCs
are uniformly old. Thus, we chose to use a constant $\Upsilon_z\equiv
1.45 \, \gdiff{M}_\odot / L_\odot$ for all galaxies.

There is an observed correlation between $r_{h}$ and $(g-z)$, in the
sense that metal-rich (red) GCs are found to be $\approx 17 \%$
smaller than their metal-poor (blue) counterparts
\citep[e.g.,][]{JCB+2005}. This can be a consequence of
mass-segregation combined with the metallicity dependence of stellar
lifetimes under the assumption that the average half-{\it mass} radius
of GCs does {\it not} depend on metallicity \citep{J2004}. The
observations of the size difference between metal-rich and metal-poor
GCs in Virgo are consistent with this explanation
\citep{JCB+2005}. Thus, it is possible that the smaller half-light
radius for metal-rich GCs do not reflect a decrease in the
corresponding half-mass radius $r_{h,M}$.

\begin{figure*}
\center
\includegraphics[width=0.75\textwidth]{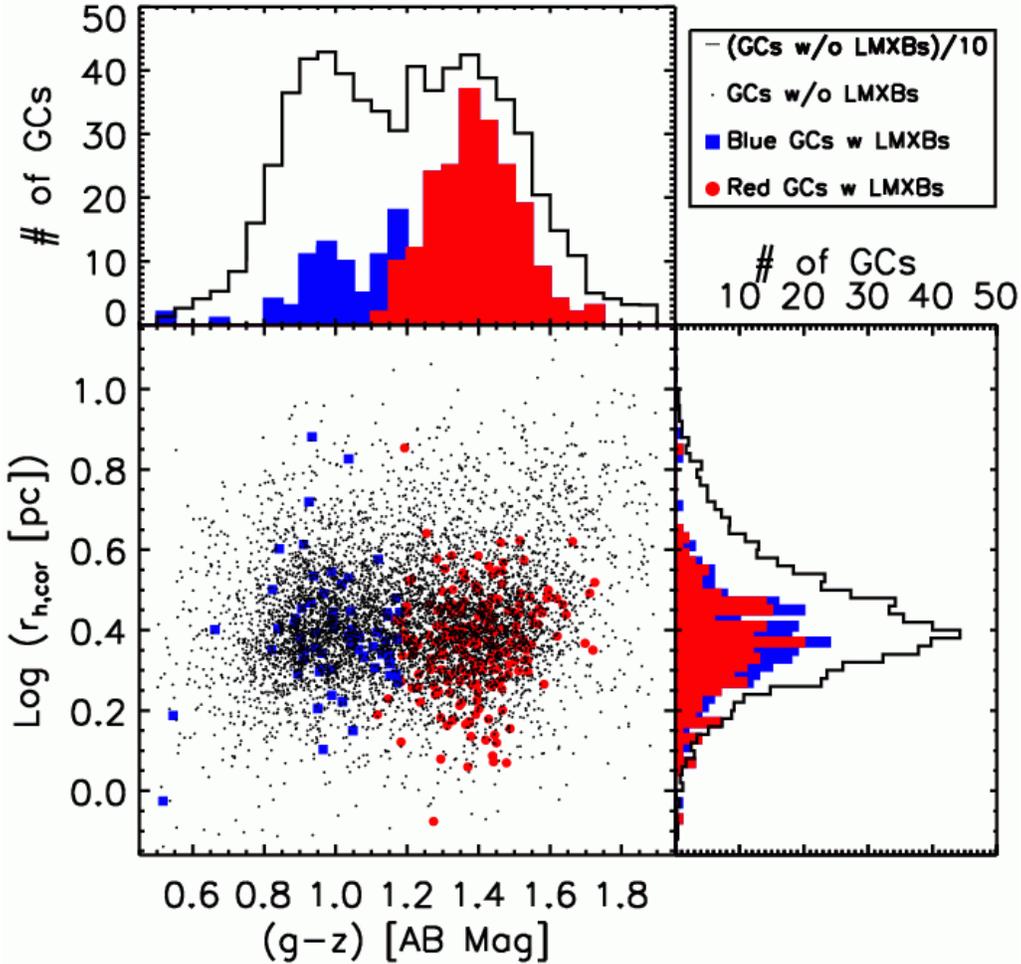}
\caption{Estimated GC half-light radii ($r_{h, {\rm
cor}}$) versus GC colors ($g-z$), with integrated histograms of the
properties above and to the right. The symbols follow those of
Figure~\ref{fig:acsvcs_gclmxb_cmd}.  There is no color dependence of
$r_{h, {\rm cor}}$, indicating we have successfully removed the color
dependence of half-light radius $r_{h}$. GCs that are redder and
smaller in extent are more likely to contain
LMXBs.\label{fig:acsvcs_gclmxb_color_size_cor}}
\end{figure*}

Since $r_{h,M}$ is more physically relevant in discussion of Galactic
dynamics than $r_{h}$, we calculated a ``corrected'' half-light radius
as
\begin{equation}
r_{h, {\rm cor}} = r_{h} \, 10^{0.17 [ (g - z) - 1.2 ]} = r_{h,M}\,
  ,
\label{eq:rhcor}
\end{equation}
following \citet{JCB+2005}, and assume in what follows that this is
the GC half-mass radius. Note that while this definition takes care of
a color dependence in the half-light radii that might not reflect a
corresponding dependence in the half-mass radii, $r_{h, {\rm cor}}$
can still differ from the actual half-mass radius by a constant factor
\citep[Figure 2 in][]{J2004}. We display the $r_{h, {\rm cor}}$ of
the GCs in our sample in Figure
\ref{fig:acsvcs_gclmxb_color_size_cor}.

Since \citetalias{BIS+2006} suggested that LMXBs occurred in GCs with
ages greater than five times the relaxation time at the half-mass
radius, $t_{h, {\rm relax}}$, we adopted the \citet{H1996} corrections
to \citet{D1993} equation (11) and used our corrected half-mass radius
estimate for calculating $t_{h, {\rm relax, cor}}$:
\begin{equation}
  t_{h, {\rm relax, cor}} = \frac{2.055 \times 10^{6}}
                                 {\ln(0.4 N_*)} \, 
                            \left( \frac{\langle m_* \rangle}
                                        {\gdiff{M}_\odot}
                            \right)^{-1} \,
                            \left( \frac{\gdiff{M}}
                                        {\gdiff{M}_\odot}
                            \right)^{1/2} \, 
                            \left( \frac{r_{h,M}}
                                        {1 {\rm \, pc}}
                            \right)^{3/2} \,
                            {\rm yr} \,
  ,
\end{equation}
where the average stellar mass, $\langle m_* \rangle$, is taken to be
$\frac{1}{3} \, \gdiff{M}_{\odot}$ and $N_* = \gdiff{M}/\langle m_*
\rangle$. \citetalias{BIS+2006} suggested that the relaxation time be combined with
the number of stars in the globular cluster to give a ``stellar
interaction rate'' of
\begin{equation}
  S = \frac{N_*}
           {t_{h, {\rm relax, cor}}} \,
      {\rm yr}^{-1}
    \propto
      \gdiff{M}^{1/2} \,
      r_{h,M}^{-3/2} \,
  .
\end{equation}

\begin{figure*}
\center
\includegraphics[width=0.75\textwidth]{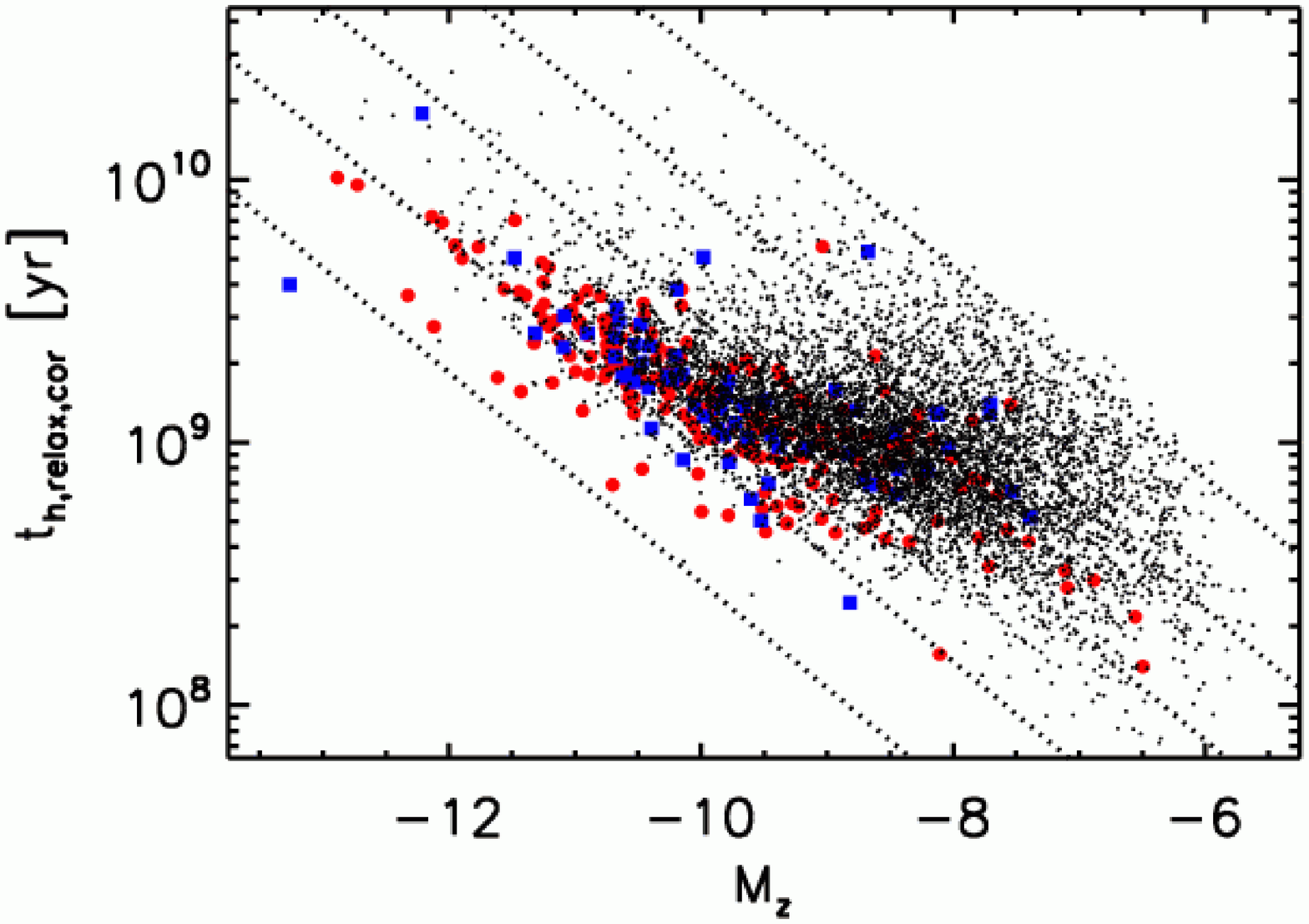}
\caption{GC relaxation time ($t_{h, {\rm relax,
cor}}$) at half-mass versus GC magnitude ($M_z$), where the
symbols are the same as in Figure~\ref{fig:acsvcs_gclmxb_cmd}. Overlaid
are lines of constant stellar crossing rate ($S = N_* /t_{h, {\rm
relax, cor}}$) that decrease by 0.5 dex starting at $\log (S \, [{\rm
yr}^{-1}]) = -2.5$ (lower left). The Galactic version of this plot for
$V$-band is displayed in \citetalias{BIS+2006}, Figure 5.  GCs with
large relaxation timescales can host LMXBs, in contrast to the result
found in the much smaller sample of Galactic GCs
\citepalias{BIS+2006}.
\label{fig:acsvcs_gclmxb_magnitude_time}}
\end{figure*}

For a system in virial equilibrium, the relaxation time is
approximately $ ( 0.1 N_* / \ln N_* ) \, t_{\rm cross}$, where $
t_{\rm cross}$ is the stellar crossing time (roughly the orbital
period of a star within the GC). Thus, $ S \approx 10 \ln ( N_* ) /
t_{\rm cross}$; except for a nearly constant factor, it is just the
inverse of the orbital period. We will refer to $S$ as the ``stellar
crossing rate.'' In Figure~\ref{fig:acsvcs_gclmxb_magnitude_time}, we
plot $t_{h, {\rm relax, cor}}$ versus $M_z$ for the GCs in our sample,
overlaying lines of constant $S$ for comparison with
\citetalias{BIS+2006}, Figure 5.

\begin{figure*}
\center
\includegraphics[width=0.75\textwidth]{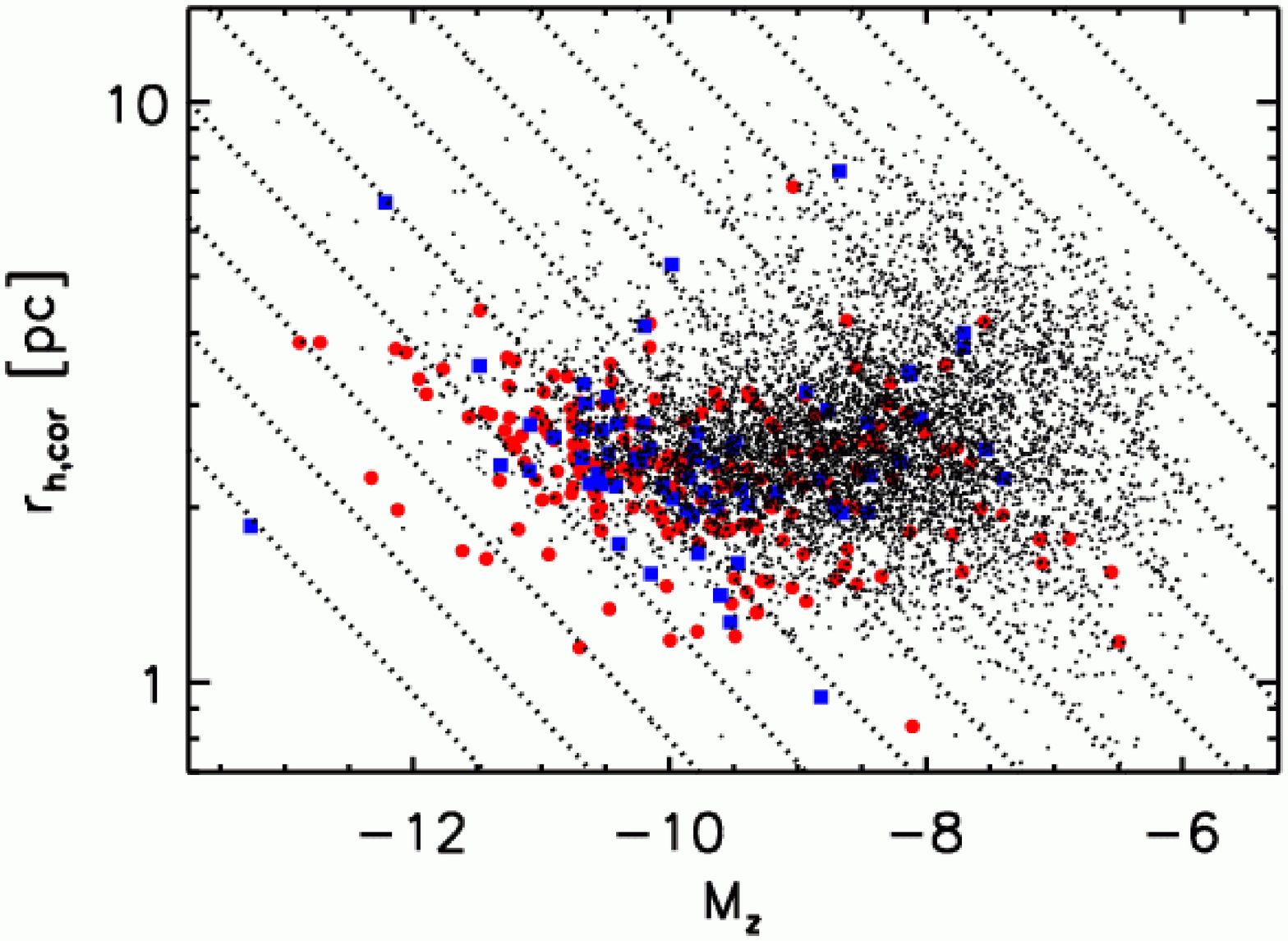}
\caption{Estimated GC size ($r_{h, {\rm cor}}$) versus
GC magnitude ($M_z$), where the symbols are the same as in
Figure~\ref{fig:acsvcs_gclmxb_cmd}.  Overlaid are lines of constant
encounter rate that decrease by 0.5 dex starting at $\log
\Gamma_h = 9.0$ (lower left). The Galactic version of this plot for
$V$-band is displayed in \citetalias{BIS+2006}, Figure 3. GCs with
larger encounter rates are more likely to contain LMXBs.
\label{fig:acsvcs_gclmxb_magnitude_size}}
\end{figure*}

The dynamical formation model of LMXBs in GCs is the leading
explanation for the larger efficiency of LMXB production in GCs
compared to the fields of galaxies. In this model, the binary is
either formed by tidal capture or through exchange interactions
between a NS/BH and an existing binary. In both cases, the
encounter rate ($\Gamma$) is thought to depend on the properties of
GCs at their cores; $\Gamma \propto \rho_0^{3/2} r_{c}^{2}$, where
$\rho_0$ is the central density, $r_c$ is the core radius, and the virial
theorem has been assumed to connect core velocities to densities and
radii \citep[see, e.g.,][]{V2003}.
Although these parameters are not directly measurable at the
distance of the Virgo cluster, we can create a proxy,
\begin{equation}
  \Gamma_{h} \equiv \left( \frac{\gdiff{M}}
                                {2\pi \gdiff{M}_\odot}
                    \right)^{3/2} \,
                    \left( \frac{r_{h,M}}
                                {1 \, {\rm pc}}
                    \right)^{-5/2} \,
  ,
\end{equation}
where we have assumed that the relation between core structural
parameters and the parameters at a half-mass radius (dictated by the
concentration, $c$) is the same for all GCs. The latter assumption is
not necessarily true, e.g. there is an observed trend in Galactic GCs
for the concentration to increase with GC mass \citep{M2000}. In any
case, we define $\Gamma_h$ to be our tracer for encounter rates under
the assumptions stated; our results on the dependence of the expected
number of GC LMXBs on $\Gamma_h$ can be easily compared with
theoretical predictions based on any alternate assumptions. We note
that if GCs with higher concentrations are more likely to contain
LMXBs, then the $\Gamma_{h}$ we calculate will underpredict the
encounter rate for GCs with LMXBs. We plot $r_{h, {\rm cor}}$ versus
$M_z$ for GCs in our sample, overlaying lines of constant
$\Gamma_{h}$, in Figure~\ref{fig:acsvcs_gclmxb_magnitude_size} for
comparison with \citetalias{BIS+2006}, Figure 3.

\citetalias{JCF+2004} used measured King models concentrations $c$ to
estimate $\Gamma$ directly. As the measured $c$ are rather uncertain
for most sources, we prefer to use $\Gamma_h$ in this work. We note
though that the results of \citetalias{JCF+2004} are robust with
respect to uncertainties in the measured $c$; indeed all conclusions
remain unchanged if a single $c\equiv 1.5$ is assumed for all GCs. In
other words, their conclusions remain the same when using $\Gamma_h$
rather than $\Gamma$.

\subsection{Matching LMXBs And GCs}

To determine the relative astrometry between {\it CXO} and {\it HST}
observations for each galaxy, we convolved the offset positions in RA
and Dec between all X-ray and optical positions with a $\sigma =
0\farcs5$ 2-dimensional Gaussian, which approximates the PSF of {\it
CXO}, to create a cross-correlated image. The maximum
cross-correlation was used to determine the astrometric offset. After
correcting for this offset, X-ray sources within $1\arcsec$ of optical
sources were considered to be matched. We chose $1\arcsec$ as a
balance between accurate identification of the X-ray source astrometry
(the astrometric accuracy of {\it CXO} in the small count regime is
$\sim 0.3$--$0.5\arcsec$) and the increased number of falsely
identified sources with larger match radii.

If an X-ray source was matched to both non-GC optical sources (e.g.,
background galaxies) and GCs, we excluded the X-ray source and GCs
from our analysis. All remaining 288 X-ray detections in GCs were
considered to be GC-LMXBs. For the 18 GC-LMXBs that were within
$1\arcsec$ of multiple GCs, we could not determine which GC contained
the LMXB; however, if the multiple GCs all belonged to the same
sub-population (blue-GC vs.\ red-GC), we could count the GC-LMXB as
belonging to that sub-population. We indicate the number of GCs with
LMXBs, and separately list the numbers of LMXBs in blue-GCs and
red-GCs, in the eighth, ninth, and tenth columns of
Table~\ref{tab:acsvcs_gclmxb_comb_sample}. In parenthesis, we indicate
the number of GC-LMXBs matched to multiple GCs. We also indicate the
corresponding number of GCs in the brackets of columns four, six, and
seven.

We separated the GCs into two sub-populations, those that clearly
contained an LMXB in our Detected sample (indicated by a subscripted
$X$), and those that clearly did not (indicated by a subscripted
$nX$). We found 270 GCs that contained an LMXB at our detection levels
and 6,488 GCs that clearly did not. We display the properties of these
GCs in Figures
\ref{fig:acsvcs_gclmxb_cmd}--\ref{fig:acsvcs_gclmxb_magnitude_time},
indicating blue-GCs with LMXBs using filled blue squares and red-GCs
with LMXBs using filled red circles. Of the 270 GCs clearly containing
an LMXB, 160 are in the SNR sample, and 61 are in the Complete sample.

We randomized the position angles of the GCs around the centers of
their host galaxies to estimate the number of LMXBs falsely associated
with GCs. We adopted two different methods: either we kept the
galactocentric distance of the GC fixed, or we assumed that the GC lay
on the same elliptical isophote on which it was originally
detected. We reassessed the matches with the randomized GC catalogs,
and used the number of resulting matches to predict the percentage of
non-GC X-ray sources falsely matched to a GC, $p_{\rm false}$. Since
there was typically little difference between the two methods, we
averaged the two. Given the typical values of these numbers
($3.3$--$14.8\%$), it is unlikely that any of the GC-LMXBs matched to
multiple GCs are not actually associated with a GC; however, some
fraction of the GC-LMXBs matched to a single GC will be false
matches. The predicted number of falsely matched sources is $p_{\rm
false}/(1-p_{\rm false})$ times the number of observed non-GC X-ray
sources. The term in the denominator accounts for the falsely matched
sources removed from the non-GC X-ray sample. The expected number of
false matches is displayed in the eleventh column of
Table~\ref{tab:acsvcs_gclmxb_comb_sample}. The large GC density in
NGC~4486 leads to a correspondingly larger number of false matches;
however, the number of false matches still make up less than $15\%$ of
the identified GC-LMXBs. With the number of false GC-LMXBs and total
number of GCs, we estimate the expected number of false matches per
GC, $\lambda_{f}$, for each galaxy, which range from about
$2\times10^{-3}$ to $7\times10^{-3}$.

\section{Properties of GCs with and without LMXBs}
\label{sec:properties}

In what follows we compare the properties of GCs that harbor an LMXB
with those of GCs that do not harbor one. We probe in turn the
distributions of luminosity (mass), $(g-z)$ color (metallicity), size
($r_h$), galactocentric distance, half-mass relaxation time $t_{ h,\rm
rel}$ and dynamical rates $S$. By revealing which properties drive the
presence of an LMXB in GCs -- or which ones are irrelevant -- this
exercise can shed light on the formation process of LMXBs in GCs.

\begin{figure*}
\center
\includegraphics[angle=90,width=0.75\textwidth]{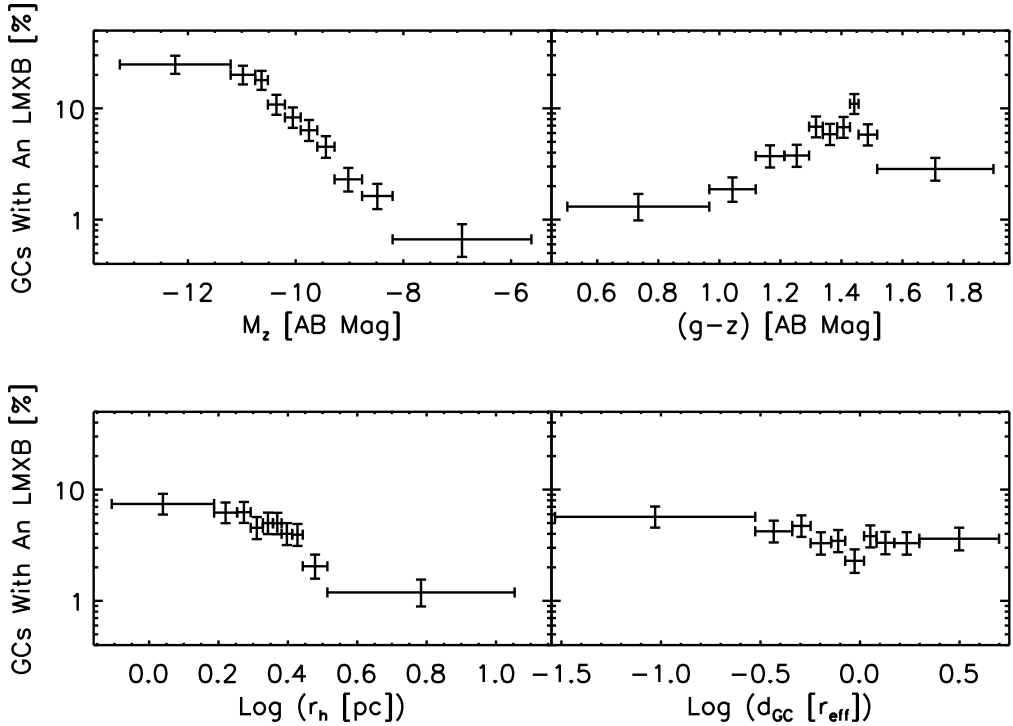}
\caption{The percentage of GCs containing an LMXB as a function of
GC magnitude ($M_z$), color ($g-z$), half-mass radius
($r_{h, {\rm cor}}$), and galactocentric distance ($d_{\rm GC}$). GCs
that are brighter, are redder, and have smaller half-mass radii are
more likely to contain LMXBs. The uncertainties are at the 1$\sigma$
confidence level. The bins were set by requiring 27 GCs with LMXBs
per bin.
\label{fig:acsvcs_gclmxb_eff}} 
\end{figure*}

We employ three methods to compare the properties of GCs with and
without LMXBs (When comparing the properties of the SNR sample and
Complete sample to GCs without LMXBs, we did not add the GCs which had
LMXBs that did not fall in the sample into the sample of GCs without
LMXBs.) First, we use binned histograms of various properties to
display the qualitative differences. These histograms are displayed in
Figures~\ref{fig:acsvcs_gclmxb_cmd}--\ref{fig:acsvcs_gclmxb_magnitude_time}. Second,
we calculate the median values of the GC properties and use the
non-parametric Wilcoxon rank-sum test \citep[$\sigma_{\rm WRS}$,
equivalent to the Mann-Whitney rank-sum test;][]{MW1947} to quantify
the differences. We additionally use the Wilcoxon test to compare the
properties of GCs with LMXBs in the Complete sample and GCs with LMXBs
fainter than the Complete sample limit. Finally, we bin the GCs by the
different parameters available and look for non-uniform probabilities
of GCs containing an LMXB. The results of the latter exercise are
shown in Figure~\ref{fig:acsvcs_gclmxb_eff}. The errors on the
fractions in Figure~\ref{fig:acsvcs_gclmxb_eff} are calculated
assuming Poisson statistics (i.e., strict 1$\sigma$ confidence
intervals were calculated as opposed to using the $\sqrt{N}$
approximation). The fractions in Figure~\ref{fig:acsvcs_gclmxb_eff}
were corrected for the rate of false matches, $\lambda_{f}$.

\subsection{Luminosity and Mass}

Prior observations have revealed that more luminous GCs appear to
preferentially host LMXBs
\citep[e.g.,][]{ALM2001,KMZ2002,SKI+2003,JCF+2004}. Figure
\ref{fig:acsvcs_gclmxb_cmd} confirms that LMXBs are found more often
in brighter GCs. The median $M_z$ is $-8.5$ for GCs without LMXBs, and
$-9.9$ for GCs with LMXBs (all samples), which correspond to Wilcoxon
rank-sum differences of 15.6$\sigma$ (Detected sample), 11.8$\sigma$
(SNR sample), and 6.6$\sigma$ (Complete sample). We find no
significant difference ($\sigma_{\rm WRS} = 0.9$) in the optical
luminosities of GCs in the Complete sample and of GCs with fainter
LMXBs.

The median mass of a GC with an LMXB is 3.6 times that of a GC without
an LMXB. The upper left panel of Figure~\ref{fig:acsvcs_gclmxb_eff}
suggests a power-law dependence of the probability of a GC containing
an LMXB on the mass of the GC.

\subsection{Color}

Prior observations also have revealed that redder GCs appear to
preferentially host LMXBs
\citep[e.g.,][]{KMZ+2003,SKI+2003,JCF+2004}. Our larger sample clearly
confirms this (see Figure~\ref{fig:acsvcs_gclmxb_cmd}). The median
$(g-z)$ color is 1.20 for GCs without LMXBs, and 1.34 (Detected
sample), 1.35 (SNR sample), and 1.38 (Complete sample) for GCs with
LMXBs, which correspond to Wilcoxon rank-sum differences of
6.6$\sigma$, 5.8$\sigma$, and 3.7$\sigma$. There were no significant
differences ($\sigma_{\rm WRS} =1.2$) in the colors of GCs in the
Complete sample and GCs with fainter LMXBs.

\begin{figure*}
\center
\includegraphics[width=0.75\textwidth]{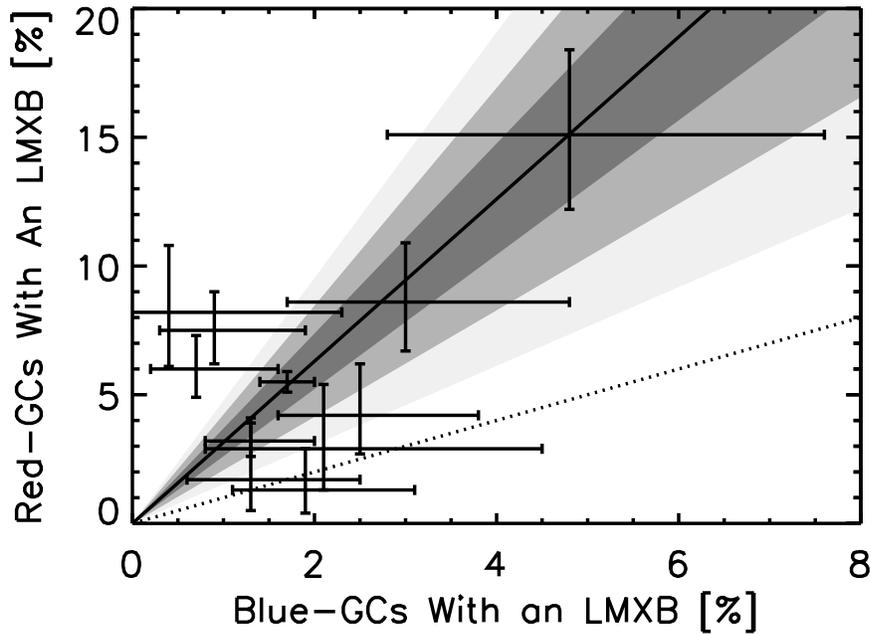}
\caption{The percentage of red-GCs containing an LMXB versus the
percentage of blue-GCs containing an LMXB for the Detected sample.
The data points with 1$\sigma$ confidence limits are the different
galaxies in our sample. The dotted line indicates the cumulative
relation that red-GCs contain as many LMXBs as blue-GCs; the data clearly
do not support this relation. The solid line indicates the cumulative
relation that red-GCs contain $3.15\pm0.54$ times as many LMXBs as
blue-GCs. The greyscale indicates the
area covered by 1, 2, and 3$\sigma$ off of the best fit.
\label{fig:acsvcs_gclmxb_rb_eff}} 
\end{figure*}

Previous works have often compared the fraction of blue-GCs with LMXBs
to the fraction of red-GCs with LMXBs, and found that red-GCs are
$\sim3$ times more likely to contain LMXBs that their blue
counterparts. In Figure~\ref{fig:acsvcs_gclmxb_rb_eff}, we compare
these fractions computed using the Detected sample. The cumulative
sample suggests that red-GCs are $3.15\pm0.54$ times more likely to
have LMXBs than blue-GCs; however, there is considerable scatter in
this fraction between galaxies.

The upper right panel of Figure~\ref{fig:acsvcs_gclmxb_eff} suggests
that the probability that a GC contains an LMXB depends exponentially
on the color. Based on the observed color distribution bimodality,
some GC formation scenarios posit that the red and blue GC
subpopulations have different formation histories \citep[see,
e.g.,][]{WCM+2004}. If the presence of an LMXB in a GC was somehow
mainly determined by this history, it is conceivable that the
probability of a GC holding an LMXB was different for each
subpopulation but independent of color within each subpopulation. If
this dependence were due only to a difference between the red and blue
GC populations, but were independent of the color of a GC within the
two populations, this would appear as a step-function relation in the
upper right panel of Figure~\ref{fig:acsvcs_gclmxb_eff}, except for
the third bluest color bin. This bin contains all but two of the
different galactic division points between red-GCs and blue-GCs. As
such, its value would be intermediate between the two
efficiencies. Since both pictures, i.e., a continuous exponential
dependence and a step-function, have two constraints, we can compare
their $\chi^2$ fits to test which interpretation is more consistent
with the data. For these fits, we always excluded the third bluest
color bin and tested the effect of excluding zero, one, or two of the
reddest color bins. Since the step-function model always had a higher
$\chi^2$ compared to the exponential function model ($\chi^2_{\rm
step}/{\rm dof} = 12.8/6, 12.8/5, {\rm \, and \, } 6.2/4$ compared to
$\chi^2_{\rm exp}/{\rm dof} = 9.7/6, 6.0/5, {\rm \, and \, } 3.7/4$,
respectively), we do not believe that the probability that a GC
contains an LMXB depends on a difference between the red and blue GC
populations that is independent of the color of a GC within the two
populations. Since color is roughly a logarithm of metallicity, we
prefer interpreting the upper right panel of
Figure~\ref{fig:acsvcs_gclmxb_eff} as an indication that the
probability a GC contains an LMXB has a continuous power-law
dependence on GC metallicity.

\subsection{Size}

Although there were indirect indications that the size of GCs affect
the formation or evolution of LMXBs in \citetalias{JCF+2004}, we
present here the first explicit evidence of this. We find that LMXBs
are found more often in GCs that have smaller half-light radii (Figure
\ref{fig:acsvcs_gclmxb_size_dist}). Because the half-light radii of
GCs are uncorrelated with their mass \citep{M2000,JCB+2005}, this
implies that LMXBs are found more often in GCs that are {\it
denser}. The median $r_{h}$ values are $2.6 {\rm \, pc}$ for GCs
without LMXBs, and $2.3 {\rm \, pc}$ (Detected sample), $2.3 {\rm \,
pc}$ (SNR sample), and $2.2 {\rm \, pc}$ (Complete sample) for GCs
with LMXBs, which correspond to Wilcoxon rank-sum differences of
8.1$\sigma$, 7.4$\sigma$, and 4.9$\sigma$. Although the differences
in the medians are $\lesssim 15\%$, the distributions are found to be
significantly different for two reasons. First, with the number of GCs
in our samples, the Wilcoxon test is sensitive to 10\%, 14\%, and 22\%
differences in the medians, respectively, at the 3$\sigma$
level. Second, the Wilcoxon test samples the entire distribution, not
just the medians. For example, the values of the half-light radii that
contain 90\% of the GCs are $4.3 {\rm \, pc}$ for GCs without LMXBs,
and $3.3 {\rm \, pc}$ (Detected sample), $3.0 {\rm \, pc}$ (SNR
sample), and $3.0 {\rm \, pc}$ (Complete sample) for GCs with
LMXBs. Kolmogorov-Smirnov tests indicate that the probability the
distributions of half-light radii are drawn from the same sample are
$5.7\times 10^{-13}$ (Detected sample), $8.1\times 10^{-11}$ (SNR
sample), and $9.6\times 10^{-6}$ (Complete sample), consistent with
the Wilcoxon tests. Once again, we found no significant differences
($\sigma_{\rm WRS} = 1.2$) in the half-light radii of GCs in the
``Complete sample'' and GCs with fainter LMXBs.

Given that redder GCs are more likely to contain LMXBs and have
smaller half-light radii, we also compared the half-{\it mass} radii
(see eq.~[\ref{eq:rhcor}]) of GCs with and without LMXBs
(Figure~\ref{fig:acsvcs_gclmxb_color_size_cor}). Our adopted half-mass
radii should be, by definition, nearly insensitive to color
variations. The median $r_{h, {\rm cor}}$ of GCs with LMXBs were only
$0.1 {\rm \, pc}$ more than their half-light radii. The distributions
of half-mass radii of GCs with and without LMXBs are different at the
5.9$\sigma$ (Detected sample), 5.5$\sigma$ (SNR sample), and
3.7$\sigma$ (Complete sample) levels according to the Wilcoxon
rank-sum test. Thus, GCs that are smaller in extent remain more likely
to contain LMXBs, even if the half-mass radii are used to measure the
size.

The lower left panel of Figure~\ref{fig:acsvcs_gclmxb_eff} suggests
that the probability that a GC has an LMXB may decrease roughly as a
power-law of the half-light radius. There is a similar, but slightly
flatter dependence on half-mass radius.

\subsection{Galactocentric Distance}

\citet{KKF+2006} present a study of LMXBs in six early-type galaxies (four
overlap with our study). Combining {\it HST}-WFPC2 and ground-based observations
they identify 285 LMXBs with GCs. They find that LMXBs are more likely (at the
$6.6 \sigma$ significance level) to be found in GCs in the inner $0.5 d_{\rm
eff}$ compared to an annulus between one-half and one times the isophotal $B=25
{\rm \, mag \, arcsec}^{-2}$ radius; 44 GC-LMXBs out of 1004 GCs are found in
the inner region versus 69 GC-LMXBs out of 2908 GCs in the outer). The suggest
their results indicate that galactocentric distance plays a critical role and
that GCs may have more compact cores near the galactic center, and thus higher
encounter rates.

The effect of galactocentric distance on the probability that a GC has
a LMXB is less clear in our data.
For comparison with their result, we examine the same inner annulus and $1.5
d_{\rm eff} < d< 3.0 d_{\rm eff}$, which corresponds to approximately the same
outer annulus. We find 66 GC-LMXBs out of 1211 GCs versus 44 GC-LMXBs out of
1203 GCs. Although the inner region is numerically more likely to have GCs
containing LMXBs in our sample, this result is not statistically significant
($1.7 \sigma$). Comparing the inner regions of the two samples, where their
study relies heavily on {\it HST} identifications, are clearly consistent. There
is a large discrepancy in the outer region. With ground-based (and to a much
lesser extent {\it HST}-WFPC2 observations), it is more likely that
contamination of GCs with unrelated objects occurs when compared with our {\it
HST}-ACS observations. Although \citet{KKF+2006} attempt to account for this
contamination, underestimation of the contamination level would result in a
larger falsely identified GC population that would deflate their measured
likelihood of finding a GC with an LMXB in the outer field.

In our data,
the values of $d_{\rm GC}$ are slightly smaller for GCs with LMXBs in the
Complete sample as compared to GCs with LMXBs not in the Complete sample, but
the difference is not very significant ($\sigma_{\rm WRS} = 2.2$). This could
indicate that the values of $d_{\rm GC}$ are affected by incompleteness. Since
the incompleteness is more important in the X-ray brighter, central regions of
galaxies, and incompleteness affects the fainter LMXBs more, this effect is not
unexpected. There is no significant difference ($\sigma_{\rm WRS} = 1.6$) in the
radial distributions between the Detected sample of GCs with LMXBs and GCs
without LMXBs; however, there are slightly significant differences when using
the SNR sample ($\sigma_{\rm WRS} = 3.1$) and Complete sample ($\sigma_{\rm WRS}
= 2.8$). The lower right panel of Figure~\ref{fig:acsvcs_gclmxb_eff} suggest
that there may be a slight decrease with $d_{\rm GC}$ in the probability that a
GC contains an LMXB. However, since red-GCs tend to be closer to the galaxy
center and are more likely to contain LMXBs, it is possible that the lower
galactocentric distances of GCs with LMXBs in the SNR and Complete sample can be
completely accounted for without invoking an intrinsic difference in the
probability a GC contains an LMXB on $d_{\rm GC}$. Given the marginal, if any,
strength of the dependence on $d_{\rm GC}$, we do not consider it further in
this paper.
We have however checked that including a galactocentric distance effect does not
alter any of our scientific conclusions.

\subsection{Relaxation Time}

No Galactic GC with a relaxation timescale $\gtrsim 2.5 {\rm \, Gyr}$
contains an active LMXB \citepalias{BIS+2006}. Their Galactic sample
is limited to 141 GCs, 12 of which contain LMXBs with $L_X > 10^{35}
{\rm \, erg \, s}^{-1}$. Our sample, which does not probe as deeply in
X-ray luminosity, includes 6,758 GCs, 270 of which contain LMXBs. In
our much larger sample (Figure
\ref{fig:acsvcs_gclmxb_magnitude_time}), about 18\% of the GCs with
LMXBs have relaxation timescales larger than this. We used Monte Carlo
simulations to estimate that measurement errors only bias this result
by about 3\%. In other words, the relaxation timescale of about 15\%
of the GCs are intrinsically larger than 2.5 Gyr. The relaxation
timescale of the GCs with LMXBs in our sample are larger than those
GCs without LMXBs; the median $t_{h, {\rm relax, cor}}$ is $1.0 {\rm
\, Gyr}$ for GCs without LMXBs, and $1.5 {\rm \, Gyr}$ (Detected
sample), $1.4 {\rm \, Gyr}$ (SNR sample), and $1.3 {\rm \, Gyr}$
(Complete sample) for GCs with LMXBs. This difference is {\it
opposite} to that seen in the Galaxy, where the median $t_{h, {\rm
relax}}$ are $0.7 {\rm \, Gyr}$ for GCs with LMXBs and $1.3 {\rm \,
Gyr}$ for GCs without LMXBs \citep{H1996}. In our sample, the
relaxation timescale distributions of GCs with LMXBs are different
from GCs without LMXBs at the 6.8$\sigma$, 4.6$\sigma$, and 2.4$\sigma$
level according to the Wilcoxon rank-sum test; the equivalent test on
Galactic GCs indicates a difference of 2.0$\sigma$. We found no
significant differences ($\sigma_{\rm WRS}=1.3$) between the half-mass
relaxation timescales of GCs in the ``Complete sample'' and GCs with
fainter LMXBs.

\subsection{Dynamical Rates}

There are two dynamical rates that could affect formation or evolution
of LMXBs: the stellar crossing rate ($ S \propto \gdiff{M}^{1/2} \,
r^{-3/2})$ and the encounter rate ($\Gamma_{h} \propto \gdiff{M}^{3/2}
\, r^{-5/2}$). Since GCs that are more massive and smaller in extent
are more likely to contain LMXBs, it is clear that both of these rates
would be higher in GCs with LMXBs compared to GCs without LMXBs
(Fig.~\ref{fig:acsvcs_gclmxb_magnitude_size} and
\ref{fig:acsvcs_gclmxb_magnitude_time}). The median values of $S$ are
$0.74 {\rm \, kyr}^{-1}$ for GCs without LMXBs and
$1.8  {\rm \, kyr}^{-1}$  (Detected sample),
$2.0  {\rm \, kyr}^{-1}$ (SNR sample), and
$1.8  {\rm \, kyr}^{-1}$ (Complete sample) for GCs with LMXBs,
corresponding to 16.9$\sigma$, 13.3$\sigma$, and $7.8\sigma$ Wilcoxon
rank-sum differences. For $\Gamma_{h}$, the median values are
$6.9\times10^5$ for GCs without LMXBs and
$7.3\times10^6$ (Detected sample),
$7.9\times10^6$ (SNR sample), and
$6.4\times10^6$ (Complete sample)
for GCs with LMXBs, corresponding to 17.1$\sigma$, 13.2$\sigma$, and
$7.7\sigma$ Wilcoxon rank-sum differences. We found no significant
differences ($\sigma_{\rm WRS}=0.1$) between either of the dynamical
rates of GCs in the Complete sample and GCs with fainter LMXBs.

\section{Multi-Variable Relation Between LMXBs and GCs}

\subsection{Technique}

Since several properties appear to affect the probability that a GC
contains an LMXB, it is important to simultaneously account for these
properties. Based on Figure \ref{fig:acsvcs_gclmxb_eff} and the
relatively small percentage of GCs falsely matched to LMXBs
$\lambda_{f}$, we have assumed the expected (true) number of LMXBs in
a GC ($\lambda_{t}$) has the following dependence on GC properties:
\begin{equation}
  \lambda_{t} = A \,
                \left( \frac{\gdiff{M}}
                            {\gdiff{M}_\odot}
                \right)^{\beta} \,
                10^{\delta \, (g-z)} \,
                \left( \frac{r}
                            {1 {\rm \, pc}}
                \right)^{\epsilon} \,
  ,
  \label{eq:lambda}
\end{equation}
where we fit for the normalization ($A$) and indices ($\beta$,
$\delta$, and $\epsilon$). Here, $r$ can either be the measured
half-light radius $r_h$ or the corrected half-mass radius $r_{h, {\rm
cor}}$. Our approach differs from the fitting performed in
\citetalias{JCF+2004}; we choose to fit the expected numbers of LMXBs per
GC, as opposed to the probability a GC contains an LMXB. The
probability must saturate at unity, whereas the expected number can be
unlimited; this makes it easier to fit simple functions (e.g.,
power-laws) to the expected number and compare with Galactic
results. The two approaches give the same results when $\lambda_{t}$
is small. By including the normalization $A$ we get a direct estimate
for the number of LMXBs in GCs and we are able to probe quantitatively
the fraction of GC X-ray point sources that are expected to result
from the integrated X-ray flux of multiple sources. Another
difference is that we include a term to account for GCs falsely
matched to LMXBs, $\lambda = \lambda_{f} + \lambda_{t}$.

The expected number of LMXBs in a GC can be converted to a probability
that there are no LMXBs assuming a Poisson distribution,
\begin{equation}
  P_{nX} = \exp (- \lambda ) \,
  ,
  \label{eq:prob_no}
\end{equation}
and the probability that there is at least one LMXB,
\begin{equation}
  P_{X} = 1 - \exp (- \lambda ) \,
  .
  \label{eq:prob_yes}
\end{equation}
One can then vary the parameters to maximize the log likelihood,
\begin{equation}
  \psi = \ln \left[ \left( {{\displaystyle \prod} \atop {\scriptstyle nX}} P_{nX} \right) \,
                    \left( {{\displaystyle \prod} \atop {\scriptstyle X}}  P_{X}  \right) \right] \,
  ,
  \label{eq:log_like}
\end{equation}
where the products are taken over the lists of GCs with no LMXBs and
GCs with LMXBs. Given that we did not find a difference in the masses,
colors, or sizes of GCs with LMXBs in the Complete sample as compared
to those GCs with fainter LMXBs, we first choose to use the better
statistics provided by the Detected sample to fit the indices. Given
the widely different luminosity limits in the Detected sample, we do
not believe the normalization we fit is physically meaningful. To
derive a physically meaningful normalization, we apply our derived
exponents to the Complete sample. Since the log likelihood can be
related to $\Delta \chi^2$, ($-2 \Delta \psi = \Delta \chi^2$), we use
the change in log likelihood for one degree of freedom (dof) to
determine one-dimensional fitting errors (1$\sigma$) on each varying
parameter.

\begin{deluxetable}{lccccc}
\tablewidth{-216.73077pt}
\tablecaption{Fits of the Expected Number $\lambda_{t}$ of LMXBs per GC\label{tab:acsvcs_gclmxb_obs_fit}}
\tablehead
{
&
\multicolumn{3}{c}{$\lambda_{t} \propto \gdiff{M}^{\beta} \, 10^{\delta \, (g-z)} \, r^{\epsilon}$}\\
\cline{2-4} 
\colhead{Row}&
\colhead{$\beta$}&
\colhead{$\delta$}&
\colhead{$\epsilon$}&
\colhead{$\psi$}&
\colhead{$\Delta$ dof}
}
\startdata
\multicolumn{6}{c}{$r = r_{h}$}\\
1 & $[0]$                     & $[0]$                  & $[0]$                   &    -1131.1 & 0\\
2 & $1.027^{+0.063}_{-0.061}$ & $[0]$                  & $[0]$                   & \phn-986.0 & 1\\
3 & $[0]$                     & $0.64^{+0.12}_{-0.11}$ & $[0]$                   &    -1113.5 & 1\\
4 & $1.009^{+0.060}_{-0.059}$ & $0.77^{+0.13}_{-0.13}$ & $[0]$                   & \phn-967.2 & 2\\
5 & $[0]$                     & $[0]$                  &  $-1.50^{+0.20}_{-0.20}$&    -1101.2 & 1\\
6 & $1.237^{+0.076}_{-0.075}$ & $0.52^{+0.15}_{-0.15}$ &  $-2.22^{+0.24}_{-0.25}$& \phn-918.4 & 3\\
\multicolumn{6}{c}{$r = r_{h, {\rm cor}}$}\\
7 & $[0]$                     & $[0]$                  &  $-1.14^{+0.20}_{-0.21}$&    -1113.6 & 1\\
8 & $1.237^{+0.076}_{-0.075}$ & $0.90^{+0.15}_{-0.14}$ &  $-2.22^{+0.24}_{-0.25}$& \phn-918.4 & 3
\enddata
\end{deluxetable}

\begin{figure*}
\center
\includegraphics[angle=90,width=0.75\textwidth]{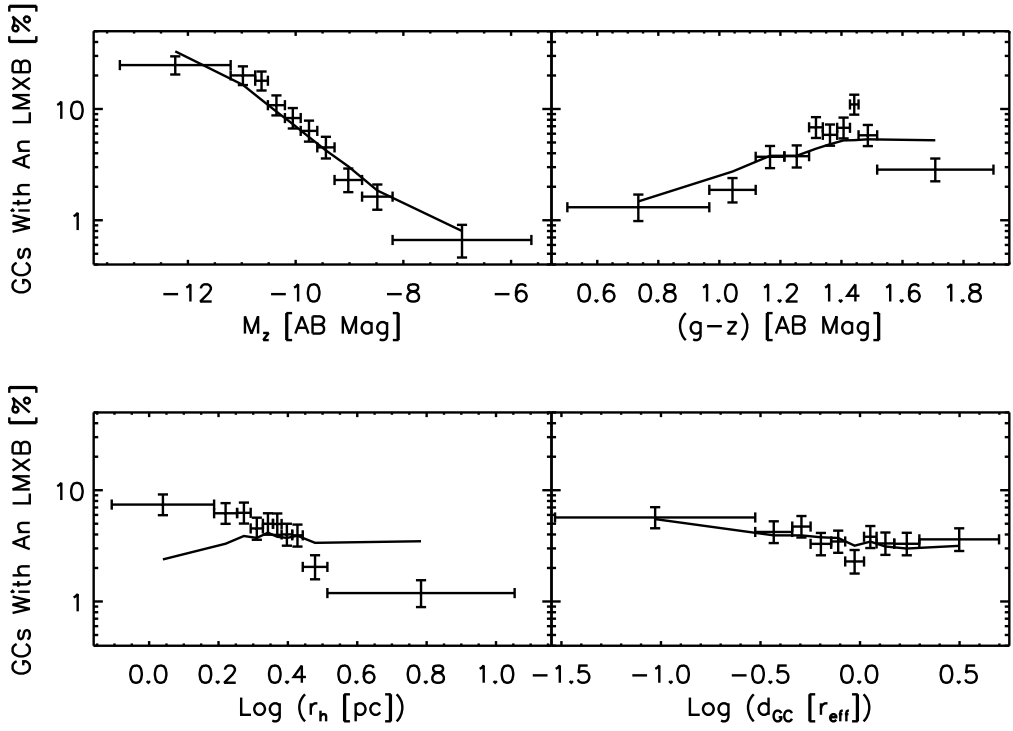}
\caption{Identical to Figure \ref{fig:acsvcs_gclmxb_eff}, except with
the results of fitting the dependence of the expected number of LMXBs
on GC mass and color (Table~\ref{tab:acsvcs_gclmxb_obs_fit}, row 4)
overlaid. Note that fitting the effects of mass and color alone does
not reproduce the observed dependence of the probability on radius or
color very well.
\label{fig:acsvcs_gclmxb_mc_fit}}
\end{figure*}

\begin{figure*}
\center
\includegraphics[angle=90,width=0.75\textwidth]{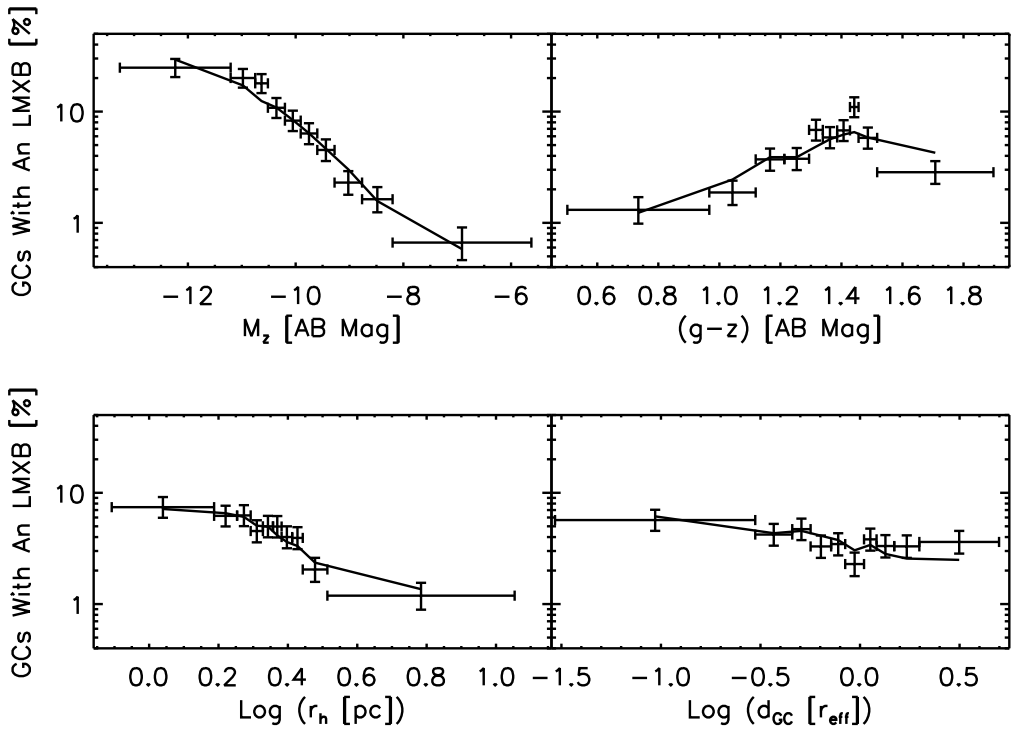}
\caption{Identical to Figure \ref{fig:acsvcs_gclmxb_eff}, except
with the results of fitting the dependence of the expected number of
LMXBs on GC mass, color, and half-mass radius
(Table~\ref{tab:acsvcs_gclmxb_obs_fit}, row 8) overlaid.
Note that simultaneously fitting the effects of GC mass, color, and half-mass
radius reproduces the observed dependence of the probability on all of the
observed parameters reasonably well.
\label{fig:acsvcs_gclmxb_mcs_fit}} 
\end{figure*}

We display the results of our fits to the Detected sample in Table
\ref{tab:acsvcs_gclmxb_obs_fit}, where bracketed numbers indicate that
an index was frozen to that number. Once a form for $\lambda_{t}$ as a
function of mass, color, and radius has been determined, we can use
this expectation to predict the fraction of GCs having an LMXB as a
function of each of the GC properties. These predictions can be
compared to the plots of the variation of this probability with
individual properties shown in Figure~\ref{fig:acsvcs_gclmxb_eff}. For
each of the binned data points in each of the panels of this Figure,
we apply the best-fit expectation calculated for each of the other
properties of the GCs in the binned data point, and then compute the
expected fraction integrated over that bin. This is compared to the
actual fraction for the bin, and the contribution to $\chi^2$ from
that bin is determined. Since we do not consider the dependence of
$\lambda_{t}$ on galactocentric distance, we did not include that
panel's contribution to $\chi^2$. Since the $\chi^2$ of that panel
ranged from 4.2--11.4 for 10 dof for all fits in
Table~\ref{tab:acsvcs_gclmxb_obs_fit}, this is justified. In this way,
we can determine whether the best-fit expectation function fits the
binned data in Figure~\ref{fig:acsvcs_gclmxb_eff} adequately. The
results, which we describe in detail in what follows, are shown in
Figures \ref{fig:acsvcs_gclmxb_mc_fit} and
\ref{fig:acsvcs_gclmxb_mcs_fit}.

\subsection{Effect of GC Mass and Color}

We first established a baseline assuming that the expected number of
LMXBs per GC has no dependence on GC properties
(Table~\ref{tab:acsvcs_gclmxb_obs_fit}, row 1); our binned $\chi^2$
was 402.4 for 29 dof ($3\times10$ data points minus one constraint for
the normalization). We next individually varied the mass and color
indices (Table~\ref{tab:acsvcs_gclmxb_obs_fit}, rows 2 and 3); mass
clearly had the largest effect ($-2\Delta \psi = 290.2$); however, its
binned $\chi^2$ was still as large as 158.6 for 28 dof. If we included
the effects of mass and color simultaneously
(Table~\ref{tab:acsvcs_gclmxb_obs_fit}, row 4), we had a much improved
fit from the mass-only fit ($-2\Delta \psi = 37.6$). We display this
fit in Figure \ref{fig:acsvcs_gclmxb_mc_fit}, which has a binned
$\chi^2$ of 106.4 for 27 dof. Although the effect of color on
$\lambda_{t}$ was included in this fit, the resulting expression does
not provide a very good fit to the dependence of the fraction on color
alone (Figure \ref{fig:acsvcs_gclmxb_mc_fit}, upper right panel). The
$\chi^2$ is 26.5 for 10 bins in this figure. In general, the
dependence of $\lambda_{t}$ on mass and color does not adequately
describe the observed number of LMXBs per GC as a function of the
observed GC properties. Some dependence on another GC property is
required.

We note that based on known correlations of GC properties we do not
expect a relation based on GC mass and color alone to be able to
reproduce any dependence on GC size. Namely, the mean $r_h$ of GCs is
observed to be independent of mass \citep{M2000,JCB+2005} and our
$r_{h, \rm cor}$ has no dependence on color. Thus, any dependence of
$\lambda_t$ on $r_{h, \rm cor}$, as observed in the lower left panel
of Figure~\ref{fig:acsvcs_gclmxb_eff}, cannot be reproduced by a
function which depends only on GC mass and color.

In the prior largest sample of GCs with LMXBs to date (6 galaxies, 98
LMXBs in 2,276 GCs), \citet{SMK+2006} found $P(LMXB) \propto
\gdiff{M}^{1.03\pm0.12} \, \gdiff{Z}^{0.25\pm0.03}$. Since they used
multiple color indices, they adopted linear color-metallicity
relations based on Milky Way GCs, ignoring scatter in the
relation. For comparison with their data, we adopt the linear
color-metallicity relation presented in \citetalias{JCF+2004} $(g-z)
\sim 0.38 \rm{[FE/H]} + 1.62$. In that case, we derive $\lambda
\propto \gdiff{M} ^{1.009^{+0.060}_{-0.059}} \gdiff{Z}^{0.29
\pm0.10}$, which is consistent with their results. However, as we
point out in the next section, the effect of GC sizes must be taken
into account. This effect alters the measured indices for mass and
color, which will in turn alter interpretations based on those
indices.

\subsection{Effect of GC Size}

There are two GC sizes we considered in our fits: the half-light
radius ($r=r_{h}$), and the half-mass radius ($r=r_{h, {\rm
cor}}$). Fitting half-light radius alone
(Table~\ref{tab:acsvcs_gclmxb_obs_fit}, row 5) leads to a better fit
than fitting color alone; however, given the correlation between color
and half-light radius, it is not clear if this is due in part to the
dependence on color or a real dependence on the cluster size. On the
other hand, only fitting the half-mass radius
(Table~\ref{tab:acsvcs_gclmxb_obs_fit}, row 7) is nearly equivalent to
only fitting the color (row 3). This is consistent with the picture
developed from the Wilcoxon tests presented in
\S~\ref{sec:properties}.

\subsection{Simultaneous Effect of GC Mass, Color, and Size}

The best fits come from simultaneously fitting the effects of mass,
color, and radius (Table~\ref{tab:acsvcs_gclmxb_obs_fit}, rows 6 and
8). Compared to just fitting the effects of mass and color, the fits
are greatly improved ($-2\Delta \psi = 97.6$), and the binned $\chi^2$
values are reasonable (20.2 for both half-light radius and half-mass
radius with 26 dof). Regardless of which size is chosen, the mass and
radius exponents are the same. The addition of the radius actually
increases the mass exponent beyond the fitting errors one derives if
radius is ignored. The color index also differs when size is
ignored. If the half-light radius is used, one finds a smaller index;
however, we believe the correlation between half-light radius and
color affects this fit, so that this does not give an accurate view of
the dependence on the size of the cluster. When the half-mass radius
is used, one finds a larger index; however, the increase is only at a
$\sim$1$\sigma$ level. The sizes of GCs play a critical role in not
only affecting the number of LMXBs per GC, but also our interpretation
of how strongly GC mass and color affect the number of LMXBs per GC.

We believe that the simultaneous fitting of GC masses, colors, and
half-mass radii provide the most physically motivated fit of the
LMXB-GC connection. For this fit (Figure
\ref{fig:acsvcs_gclmxb_mcs_fit}), we derive
$\beta = 1.237 ^{+0.076}_{-0.075}$,
$\delta= 0.90^{+0.15}_{-0.14}$, and
$\epsilon = -2.22 ^{+0.31}_{-0.36}$.
We note that by including the effect of GCs falsely matched to LMXBs
we increased the mass exponent by 13\%, increased the color index by
16\% and decreased the radius exponent by 16\%.

If we require that the expected number of LMXBs scales linearly with
the mass ($\beta \equiv 1$), the negative log-likelihood function is
increased by $-2\Delta \psi = 10.7$ for 1 more dof, and the binned
$\chi^2$ is 30.9 for 27 dof. Thus, we rule out a linear
proportionality of mass at the 99.89\% confidence limit. We discuss
the implications of this in the next subsection.

From Figure~\ref{fig:acsvcs_gclmxb_mcs_fit}, we see that the declining
probability for a GC with $(g-z)>1.4$ to contain an LMXB can be
reproduced although we assume no break in the relation between color
and $\lambda$. This suggests that the masses and half-mass radii of
GCs at these colors may be different than at bluer colors. Indeed,
the fraction of GCs with $r_h \ga 3$ pc increases systematically with
color for $(g-z)>1.4$. One possible explanation is contamination by
diffuse stellar clusters, which are redder, fainter cousins of GCs
\citep[e.g.,][]{LB2000,PCJ+2006}. Additionally, we note that the small
variations of the probability with galactocentric distance are
reproduced, even though the distance played no role in the fit.

We used the indices from our best-fit to the Detected sample to
determine the normalization for the Complete sample:
\begin{equation}
  \lambda_{t} = 8.0\times 10^{-2} \,
                \left(\frac{\gdiff{M}}
                           {10^{6} \, \gdiff{M}_{\odot}}
                \right)^{1.237} \,
                10^{0.90 \, (g-z)} \,
                \left(\frac{r_{h, {\rm cor}}}
                           {1 {\rm \, pc}}
                \right)^{-2.22} \,
  .
\end{equation}
Since our method predicts the expected number of LMXBs per GC, we can
determine the likelihood that a GC would contain multiple LMXBs. Note
that M15 in our Galaxy has two active LMXBs \citep{WA2001}; however,
the LMXBs in M15 have $L_X \sim 10^{36} {\rm \, erg \, s}^{-1}$ and
would not be detected at the distances of the galaxies in our
sample. We calculate the number of GCs containing multiple LMXBs,
$N_{\rm mult}$, as
\begin{equation}
  N_{\rm mult} = \sum_{X,nX} [ 1-(1+\lambda_{t}) e^{-\lambda_{t}} ],
\end{equation}
obtaining $N_{\rm mult} = 1.3$. Normalizing this to the number of GCs
with LMXBs in the Complete sample (61), we estimate that
$2.2^{+3.9}_{-1.7}\%$ of GCs with LMXBs actually contain multiple
LMXBs such that their combined luminosity is above $3.2\times 10^{38}
{\rm \, erg \, s}^{-1}$. This fraction should increase as the LMXB
luminosity limit of the sample decreases. For example, the same
calculation for our Detected sample with lower luminosity LMXBs gives
an expected fraction of $8.2^{+2.0}_{-1.7}\%$ of GCs with multiple
LMXBs.

\subsection{Implications of Simultaneous Fit}

For multivariate datasets, principal component analysis determines the
set of orthogonal eigenvectors, linear combinations of the original
variables, that minimize the Euclidean distances to each eigenvector
\citep[e.g., ][]{MH1987}. The normalized variances of the data
projected onto the eigenvectors, the eigenvalues, are used to order
the eigenvectors. The eigenvectors with larger eigenvalues are the
best-fit eigenvectors to the data. Often, only eigenvectors with
eigenvalues greater than unity (i.e., those eigenvectors that are as
good fits as the original variables) are retained. Principal component
analysis among the masses, colors, and radii of GCs with LMXBs found
two eigenvectors whose eigenvalues were above unity. The first
eigenvector has an eigenvalue of 1.22 and is dominated by an
approximately equal combination of mass and half-mass radius, while
the second eigenvector has an eigenvalue of 1.01 and is dominated by
color. Therefore, we break our interpretation of the fits into two
parts: an interpretation of the variation with color as a metallicity
effect, and a discussion of the combined mass and half-mass radius
variation as evidence for the dynamical formation of LMXBs in GCs.

\subsubsection{Metallicity Effect on the Presence of LMXBs}

We find a monotonic increase in the expected number of LMXBs with GC
color of the form $\lambda_{t} \propto 10^{0.90^{+0.15}_{-0.14}
(g-z)}$. Using data from NGC~4486, \citetalias{JCF+2004} found that
the probability that a GC contains an LMXB is $P_{X} \propto
10^{0.87^{+0.22}_{-0.22} (g-z)}$. For small values of the expected
number ($\lambda_{t} \ll 1$), these two quantities are nearly
identical ($P_{X} \approx \lambda_{t}$,
eqn.~[\ref{eq:prob_yes}]). Thus, our slightly different techniques are
in good agreement with respect to the fit to color variation. In
\citetalias{JCF+2004}, the color variation was directly converted into
a metallicity variation by applying the \citet{BC2003} models, which
give $(g-z) \sim (0.38\pm0.05) \log \frac{ \gdiff{Z} }{
\gdiff{Z}_\odot }$ and therefore $P_X \propto \gdiff{Z}^{0.33\pm0.1}$.
In this paper, we converted color to metallicity following equation
(2) of \citet{PJC+2006},
\begin{equation}
  \log \frac{ \gdiff{Z} }{ \gdiff{Z}_\odot } = \left\{
    \begin{array}{cc}
    -6.21 + (5.14\pm0.67) (g-z) & {\rm for \ }  g-z \le 1.05 \,
    ;\\
    -2.75 + (1.83\pm0.23) (g-z) & {\rm for \ }  g-z > 1.05 \,
    .
    \end{array}
    \right.
\end{equation}
We then replaced the $10^{(g-z) \delta}$ term from equation
\ref{eq:lambda} with $(\gdiff{Z}/\gdiff{Z}_\odot)^{\delta^\prime}$. We
find $\delta^\prime=0.391^{+0.070}_{-0.067}$, with no change to the
variation with mass or radius. The dependence on $\gdiff{Z}$ we find
here is consistent with that found in \citetalias{JCF+2004} and with
recent results on the metallicity dependence of the production of
likely quiescent LMXBs in Galactic GCs \citep{HWC+2006p}. Compared to
fitting the effect of mass, radius, and color, the mass, radius, and
metallicity fit is very slightly worse ($-2 \Delta \psi = 2.0$);
however, the binned $\chi^2$ value is still reasonable (20.3 for
half-mass radius, both with 26 dof).

There are several ideas that could explain the relation between
metallicity and the likelihood that a GC will contain an LMXB, most of
which were summarized in \citetalias{JCF+2004}. First, metal-rich
stars may have larger radii and masses compared to metal-poor stars
\citep{BPF+1995}, which would make it easier to form LMXBs. This would
increase both the number of Roche-lobe overflow systems and the number
of binary systems formed. Following \citet{MKZ2004}, we
have estimated that the number of Roche-lobe overflow systems has a
weak metallicity dependence, $\gdiff{Z}^{0.11}$. The number of NS
binaries formed by tidal capture goes as $\gdiff{Z}^{0.12}$, but may
have a lower exponent for exchange interactions. Thus, this model
predicts a $\gdiff{Z}^{\lesssim0.2}$ dependence and is inconsistent
with our fit.

\citet{I2006} suggested that the link between metallicity and outer
convective zones was responsible for the increased formation
efficiency of LMXBs in metal-rich GC. In this model, metal-poor
main-sequence stars lack an outer convective zone. This absence turns
off magnetic braking. Since magnetic braking leads to the strong
orbital shrinkage that eventually leads to mass transfer as an LMXB,
metal-poor stars fail to dynamically form many of the main-sequence/NS
binaries that can appear as bright LMXBs. Although this model is
intriguing, more work must be done to see if the removal of the outer
convective zone would predict a dependence of the expected
number of LMXBs on the GC metallicity that is consistent
with our best-fit power-law dependence.

\begin{figure*}
\center
\includegraphics[width=0.75\textwidth]{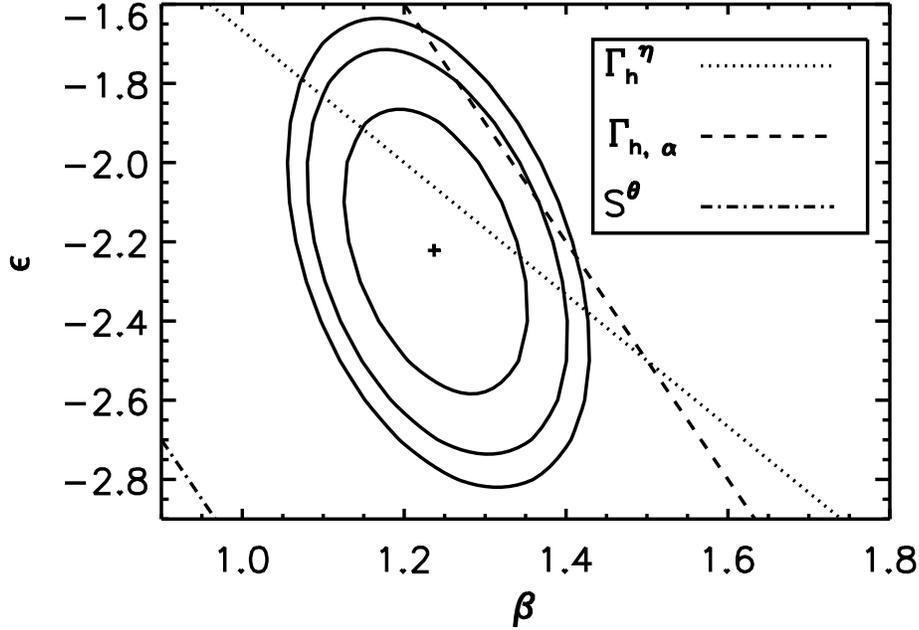}
\caption{Two-dimensional confidence intervals (at $1\sigma$, 90\%, and
$2\sigma$) of the exponents of mass ($\beta$) and radius ($\epsilon$)
in the fit to the expected number of LMXBs per GC, $\lambda \propto
\gdiff{M}^{\beta} \, 10^{\delta \, (g-z)} \, r_{h, {\rm
cor}}^{\epsilon}$. The cross indicates the best-fit values $\beta$
and $\epsilon$. The overlaid lines indicate the dependences of three
likely dynamical-formation properties that depend on mass and radius:
$\lambda_{t} \propto \Gamma_{h}^\eta$,
$\lambda_{t} \propto \Gamma_{h, \alpha} \equiv \Gamma_{h} \, \rho^\alpha$,
and
$\lambda_{t} \propto S^{\theta}$.
The power-law dependence on $\Gamma_{h}$ is the most consistent,
and a power-law dependence on $S$ can be ruled out.
\label{fig:acsvcs_gclmxb_ms_conf}}
\end{figure*}

If metal-rich GCs produce more NSs and BHs per unit mass, then the
larger number of NSs and BHs could increase the number of LMXBs that
can form. For instance, the initial mass function (IMF) could vary
with metallicity \citep{G1987}. Following \citetalias{JCF+2004}, we
assume a power-law IMF, with the number of stars with masses between
$m$ and $m + dm$ varying as $N(m) \propto m^{-x} dm$. We also assume
that the IMF slope depends on metallicity, $dx/d\mbox{[Fe/H]} \equiv
A$. The IMF metallicity dependence must be $A = -0.32
^{+0.062}_{-0.072} {\rm \, dex}^{-1}$ to account for the metallicity
dependence of the number of LMXBs per GC. As in \citetalias{JCF+2004},
the IMF dependence is in rough agreement with the analysis of LMXBs in
the Galactic and M31 GC systems, $A\sim -0.4 {\rm \, dex}^{-1}$
\citep{BPF+1995}, and IMF slope derived for Galactic GCs, $A\approx
-0.5 {\rm \, dex}^{-1}$ \citep{DPC1993}. We conclude that a mildly
metallicity dependent IMF can explain the metallicity dependence of
the number of LMXBs per GC. However, we caution that there is evidence
that massive star IMFs are metallicity independent \citep[e.g., IMFs
from OB associations;][]{MJD1995}.

Irradiation-induced winds, which would be weaker in metal-rich stars
due to more efficient metal line cooling, may affect the number of
LMXBs observed in systems that are richer in metals. \citet{MKZ2004}
presents two basic models that consider the effect of these winds. In
the first model, the wind is ejected at the escape velocity of the
donor star. At the same accretion rate (i.e., X-ray luminosity), stars
that are richer in metals will have lower mass-loss and longer
lifetimes. The rough predicted metallicity dependence for the number
of LMXBs per GC is approximately $\gdiff{Z}^{0.35}$, although we note
that in making this estimate \citet{MKZ2004} consider only the cooling
function while the strength of stellar winds due to irradiation
depends on both heating and cooling processes \citep[e.g.,][]{BS1973}.
In the second model, a less dense, higher velocity wind is ejected. In
this model, the mass-loss rates (i.e., lifetimes) of stars that are
poorer in metals and stars that are richer in metals are about the
same, but the accretion rates are not. Stars that are richer in metals
are more luminous and the predicted metallicity dependence for the
number of LMXBs per GC is roughly $\gdiff{Z}^{\sim0.7 (\beta_{\rm
lf}-1)}$, where $\beta_{\rm lf}$ is the differential X-ray luminosity
function slope. Typical values of $\beta_{\rm lf}$ are 1.8--2.2
\citep{KF2004} for a single power law; however, the X-ray luminosity
function may be consistent with a broken power law. At the higher
luminosities ($L_X > 5 \times 10^{38} {\rm \, erg \, s}^{-1}$),
$\beta_{\rm lf} = 2.8\pm0.6$. At the lower luminosities that are
characteristic of most LMXBs, $\beta_{\rm lf} = 1.8\pm 0.2$. Thus a
wide range of metallicity exponents are possible, but most are steeper
than $\gdiff{Z}^{\sim0.4}$. 
Either of these models may be consistent
with our fit; however, more detailed work developing
irradiation-induced wind models is necessary to predict the
metallicity dependence accurately.
We note that systems with stronger irradiation-induced winds are predicted to
have more absorption and that spectral analysis by X-ray colors in
\citet{KKF+2006} do not find evidence for this absorption. An ongoing similar
analysis for the sample in this paper, as well as more detailed X-ray spectral
fitting, should shed additional light on this issue.

Finally, we note that many of the above models assume that the compact
object in the LMXB binaries is a NS. However, our Complete sample
contains GC-LMXBs above $3.2\times 10^{38} {\rm \, erg \,
s}^{-1}$. Such sources are above the Eddington limit for a
hydrogen-accreting neutron star, and the compact object in the system
might be a BH. These models need to be developed to include the
effects of such binary systems.

\subsubsection{Dynamical Formation of LMXBs in GCs}

\begin{figure*}
\center
\plottwo{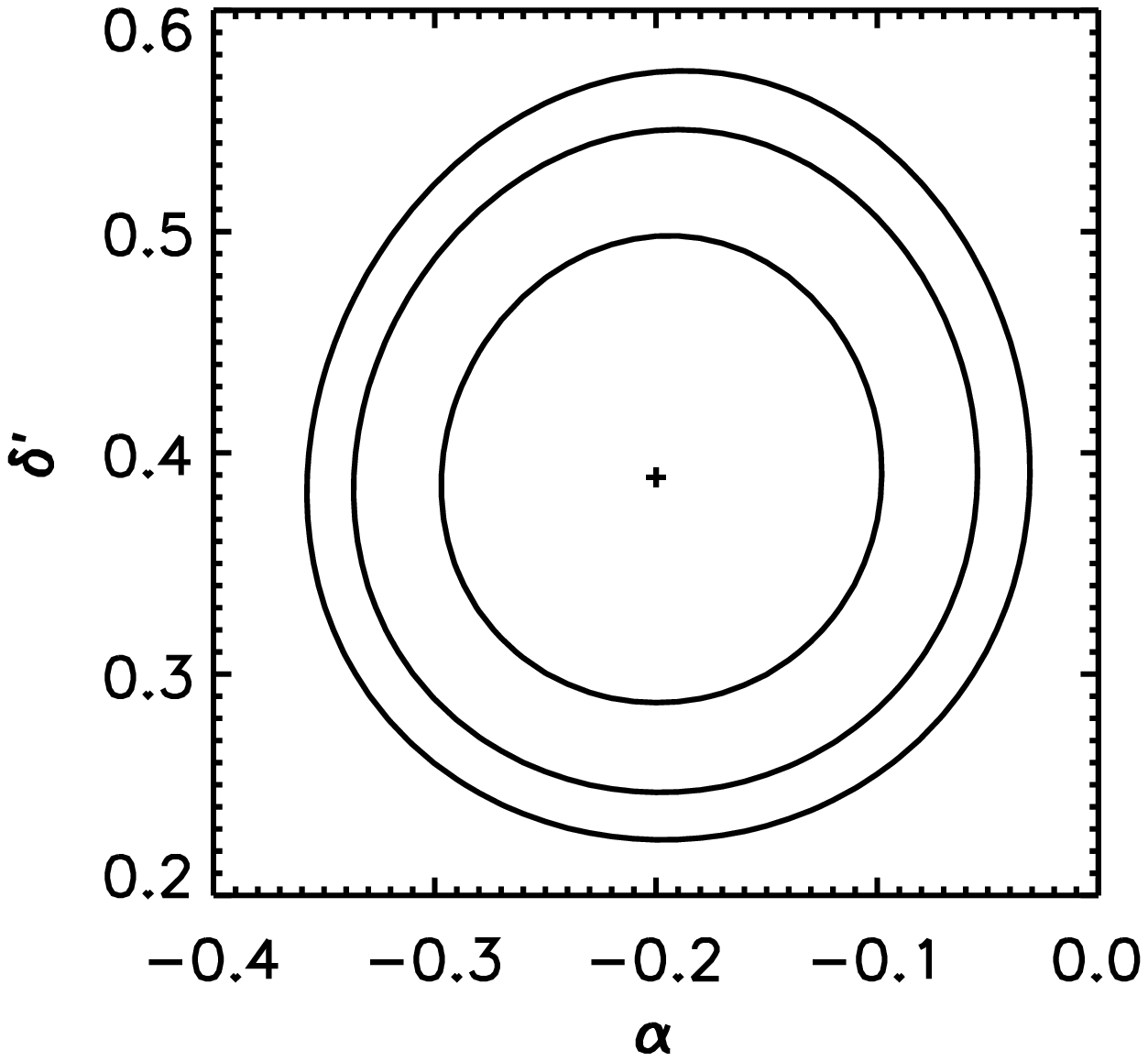}{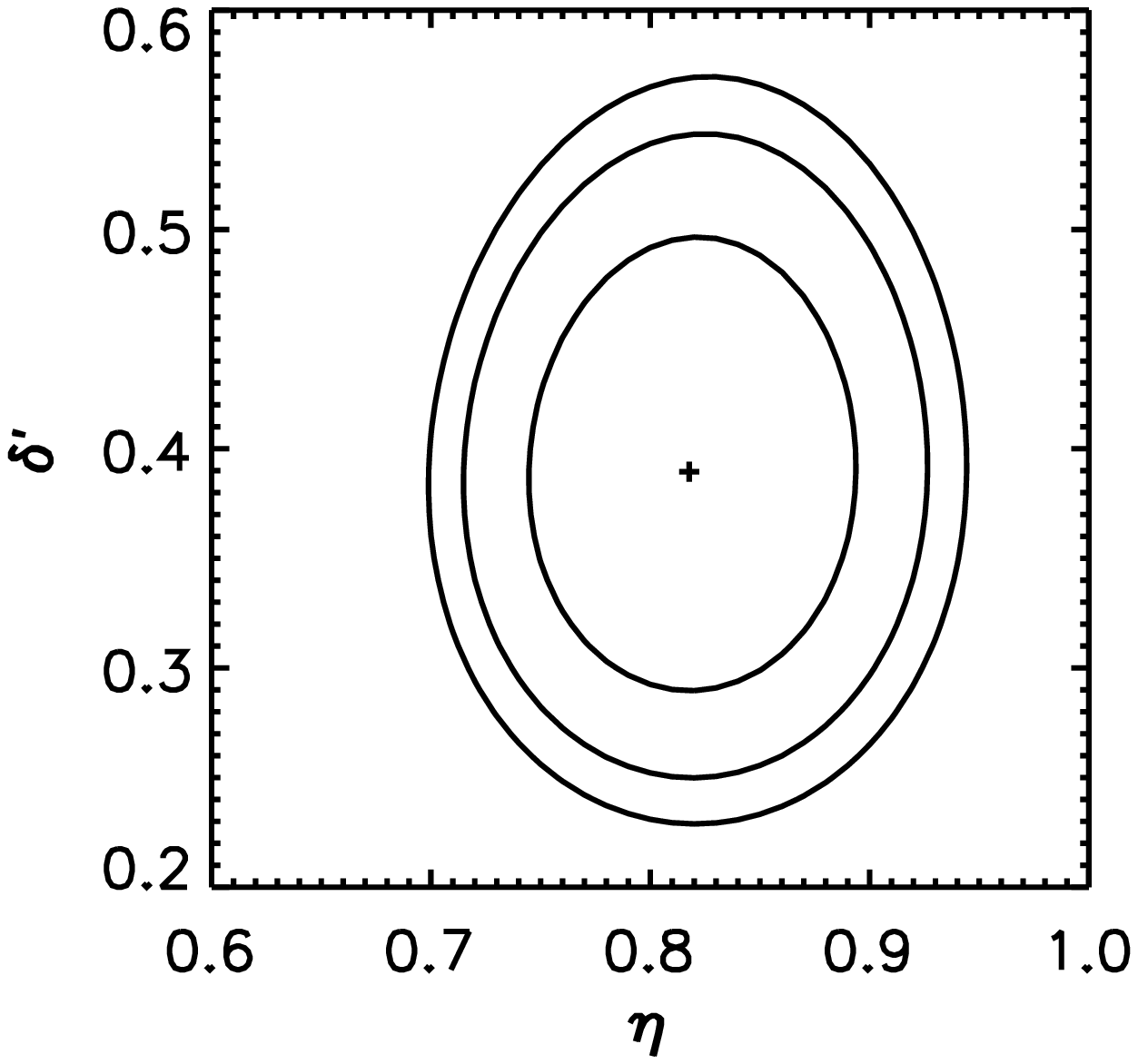}
\caption{({\it Left:}) Two-dimensional confidence intervals (at
$1\sigma$, 90\%, and $2\sigma$) of the exponents of density and
metallicity in the fit to the expected number of LMXBs per GC,
$\lambda \propto \Gamma_{h} \, ( \gdiff{M} / r_{h, {\rm cor}}^3)^\alpha \,
\left(\gdiff{Z}/\gdiff{Z}_\odot\right)^{\delta^\prime}$.
The cross indicates the best-fit values of $\alpha$ and
$\delta^\prime$. ({\it Right:}) The confidence intervals for the
exponents of the encounter rate and metallicity in the fit to
the expected number of LMXBs per GC, $\lambda \propto \Gamma_{h}^\eta
\, \left(\gdiff{Z}/\gdiff{Z}_\odot\right)^{\delta^\prime}$. The
cross indicates the best-fit values of $\delta^\prime$ and $\eta$.
\label{fig:acsvcs_gclmxb_dm_conf}}
\end{figure*}

Since the efficiency of the dynamical formation of LMXBs in GCs is
likely to depend on a combination of the GC mass and size, we plot the
two-dimensional confidence interval of the exponents for both
properties in Figure \ref{fig:acsvcs_gclmxb_ms_conf}. The effect of GC
color was also included in these fits, but is not shown. On top of the
figure, we have overlaid three lines:
$\lambda_{t} \propto S^{\theta}$
(powers of the stellar crossing rate),
$\lambda_{t} \propto \Gamma_{h}^\eta$
(powers of the encounter rate), and
$\lambda_{t} \propto \Gamma_{h, \alpha}
                   = \Gamma_{h} \rho^\alpha
                   = \gdiff{M}^{3/2+\alpha} \, r_{h, {\rm cor}}^{-5/2-3\alpha}$
(the \citetalias{JCF+2004} form of the encounter
rate\footnote{$\alpha$ in this equation is related to that in equation
(13) of \citetalias{JCF+2004}, $\alpha_{\rm VCS3}$, by $\alpha_{\rm
VCS3} = 1.5 + \alpha$}). The mass and radius exponents clearly
indicate that powers of the stellar crossing rate do not fit the data.

The \citetalias{JCF+2004} form of the encounter rate cannot be ruled
out when considering the 2-dimensional confidence interval. However,
fitting
\begin{equation}
  \lambda_{t} = A \,
                \left( \frac{\gdiff{M}}
                            {\gdiff{M}_\odot} 
                \right)^{3/2+\alpha} \,
                \left( \frac{r_{h, {\rm cor}}}
                            {1 {\rm \,  pc}}
                \right)^{-5/2-3\alpha} \,
                \left( \frac{\gdiff{Z}}
                            {\gdiff{Z}_\odot}
                \right)^{\delta^\prime} \,
  ,
\end{equation}
we find $\alpha = -0.200^{+0.066}_{-0.067}$ and
$\delta^\prime=0.389^{+0.072}_{-0.068}$ is a worse fit ($-2 \Delta
\psi = -4.4$) than fitting mass, radius, and metallicity. With only
one more dof, this fit is marginally (in)consistent at the 96.5\%
confidence limit. The slightly worse fit is also suggested by the
binned $\chi^2$ value of 33.56 for 27 dof. Any slight disagreement
must be tempered by the presence of different assumptions in the
analysis, namely the fact that \citetalias{JCF+2004} did not assume a
fixed $c$ for all GCs as we do here. We do note that the value of
$\alpha\sim-0.2$ is roughly consistent with analysis of pulsars
in Galactic GCs \citep{JKP1992}.
We display the 2-dimensional confidence interval of
$\alpha$ and $\delta^\prime$ in the left panel of Figure
\ref{fig:acsvcs_gclmxb_dm_conf}.

The best fit comes from powers of the encounter rate. When we fit
\begin{equation}
  \lambda_{t} = A \,
                \Gamma_{h}^\eta \,
                \left(\frac{\gdiff{Z}}
                           {\gdiff{Z}_\odot}
                \right)^{\delta^\prime} \,
  , 
\end{equation}
we find $\eta = 0.818^{+0.050}_{-0.049}$ and
$\delta^\prime=0.389^{+0.071}_{-0.067}$ has only a slightly worse fit
($-2 \Delta \psi = -0.6$) than fitting mass, radius, and
metallicity. Since it has one more dof, this is our preferred fit. Its
binned $\chi^2$ value was 22.2 for 27 dof.
Again, we
used the indices from our best-fit to the Detected sample to determine
the normalization for the Complete sample:
\begin{equation}
  \lambda_{t} = 4.1\times 10^{-2} \,
                \left(\frac{\Gamma_{h}}
                           {10^{7}}
                \right)^{0.818} \,
                \left(\frac{\gdiff{Z}}
                           {\gdiff{Z_{\odot}}}
                \right)^{0.389} \,
  .
\end{equation}

The dependence of
$\lambda_{t}$ on encounter rate matches the Galactic values found
for low-luminosity X-ray sources by
\citet{JV1996} of $\alpha\sim0.2$ and by 
\citet{PLA+2003} of $\lambda \propto \Gamma^{0.74\pm0.36}$.
Since low-luminosity X-ray sources represent a mixture
\citep[e.g.,][]{BPH+2004,HGL+2003,HWC+2006p,PLA+2003} of quiescent LMXBs
(qLMXBs), cataclysmic variables (CVs), and chromospherically active binaries
(CABs), such samples likely trace additional source formation and evolution
processes when compared to a sample of active LMXBs. For example,
\citet{PLA+2003} and \citet{HWC+2006p} show indications that qLMXBs are more
consistent with a linear $\Gamma$ dependence than fainter, hard X-ray sources
likely to include CVs and CABs. For qLMXBs, \citet{PLA+2003} finds $\lambda
\propto \Gamma^{1.0\pm0.5}$. To compare to qLMXBs from \citet{HWC+2006p}, we
use the \citetalias{JCF+2004} form of the encounter rate. Our $\alpha$ is
defined such that it is 1.5 less than the Alpha used in Figure 8, {\it right} of
\citealt{HWC+2006p}), and our $\delta^{\prime}$ is their Delta. These values lie
close to the $>50\%$ probability contour, closer than a linear dependence on
$\Gamma$ without any metallicity dependence. However, the $>50\%$ probability
contour value includes a large range of values both shallower (to the left of
Alpha = 1.5) and steeper (to the right of Alpha = 1.5) than a linear dependence
on $\Gamma$.
The relatively limited numbers of Galactic
qLMXBs in current samples make it difficult for these studies to constrain 
the dependence of the production rates of LMXBs in GCs.

\citetalias{JCF+2004} suggested two mechanisms
to account for our observed less than linear dependence on the encounter rate;
it might result from the competing rate of destruction of binaries, or it might
be due to hardening (shrinking) of binaries.

If both the formation and destruction rates of the close binaries per unit
stellar mass) depend on the encounter rate parameter per unit mass $\Gamma_h /
\gdiff{M}$ \citep{V2003} then in steady-state the encounter rate would cancel
and the number of LMXBs per GC would just be proportional the number of stars,
or $\lambda_t \propto \gdiff{M}$. This would be inconsistent with the
dependencies of $\lambda_t$ on mass, radius, and on $\Gamma_h$ that we have
found. However, if the destruction rate of binaries due to encounters is
comparable to but less than the destruction rate due to stellar evolution of the
binaries, then an intermediate dependence on $\Gamma_h / \gdiff{M}$
(between linear and none) might result. Detailed simulations of binary
formation, evolution, and destruction in GCs are needed to test this
hypothesis.

During exchange interactions, the orbits of wide binaries can be
reduced, particularly in denser clusters. Following
\citetalias{JCF+2004}, the binary encounter cross-section, $C$, can be
written as a combination of the cross-section for tidal capture,
$\sigma_2$, and the cross section for binary exchange interactions,
$\sigma_3 \, \zeta(\rho)$, where $\zeta(\rho)$ encapsulates the effect
of hardening as a function of density. If binary hardening is
responsible for $\alpha = -0.2$, then $C = \sigma_2 + \zeta(\rho)
\sigma_3 \propto \rho^{-0.2}$. This would yield $\zeta(\rho) = (\mu \,
\rho^{-0.2} - \sigma_2)/\sigma_3$, where $\mu$ is a constant, as a
constraint on predictions for binary hardening.

\section{Conclusions}

We have compared the masses, colors, sizes, and galactocentric
distances of 270 GCs that contain LMXBs and 6488 GCs that do not
contain LMXBs in a sample of eleven galaxies. There are clear
differences between the masses, colors, and radii of GCs in these two
samples, while the role of galactocentric distance has at most a weak
effect. There is no significant difference between the masses, colors,
or sizes of 61 GCs that contain LMXBs in the Complete sample ($L_X \ge
3.2 \times 10^{38} {\rm \, erg \, s}^{-1}$) and 209 GCs with fainter
LMXBs; thus, there is no evidence that luminous and fainter LMXBs are
affected differently by the properties of the GCs they inhabit.

We clearly show that GCs that are more massive and redder are more
likely to contain LMXBs, confirming previous findings. Although we
find that red-GCs are $3.15\pm0.54$ times more likely to have LMXBs
than blue-GCs, we show that the detailed dependence changes
continuously with GC color, rather than being a simple function of the
overall population to which the GC belongs.

With the half-mass radius, we calculate the relaxation timescale. In
contrast to Galactic GCs \citepalias{BIS+2006}, we find that a large
number of GCs that contain LMXBs ($\sim15\%$) have relaxation times
$>2.5 {\rm \, Gyr}$. Based on the results from our sample, it does not
appear necessary for GCs to survive for more than five relaxation
timescales in order to produce LMXBs.

Most notably, we find that GCs that have smaller half-light radii or
half-mass radii are more likely to contain LMXBs. This paper presents
the first clear indication that GC half-mass radius affects the
likelihood a GC will contain an LMXB, although a size dependence was
implicit in the results of \citealp{JCF+2004}.

Simultaneous fits of the dependence of the expected number of LMXBs
per GC on the GC mass, color, and radius gave
\begin{equation}
  \lambda_{t} \propto \gdiff{M}^{1.237^{+0.076}_{-0.075}} \,
                      10^{0.90^{+0.15}_{-0.14} \, (g-z)} \,
                      r_{h, {\rm cor}}^{-2.22^{+0.24}_{-0.25}} \,
  .
\end{equation}
Including the radius is important because fitting mass and color alone
does not provide an adequate fit to the expected number of LMXBs, and
the form of the dependence on mass and color changes when the effects
of the size are included. The simplest model for LMXBs, that the
number of LMXBs per GC is linearly proportionality to GC mass, can be
ruled out at the 99.89\% confidence limit.

For our Complete sample, $L_X > 3.2\times 10^{38} {\rm \, erg \,
s}^{-1}$, we find
\begin{equation}
\lambda_{t} = 8.0\times 10^{-2} \,
              \left(\frac{\gdiff{M}}
                         {10^{6} \, \gdiff{M}_{\odot}}
              \right)^{1.237} \,
              10^{0.90 \, (g-z)} \,
              \left(\frac{r_{h, {\rm cor}}}
                         {1 {\rm \, pc}}
              \right)^{-2.22} \,
.
\end{equation}
We additionally provide for the first time an expression to estimate
the expected number of multiple LMXB sources in GCs, which predicts
that $2.2^{+3.9}_{-1.7}\%$ of GCs with LMXBs actually contain multiple
LMXBs such that their combined luminosity is above $3.2\times 10^{38}
{\rm \, erg \, s}^{-1}$. Thus, we predict that most GCs with high
X-ray luminosities contain a single LMXB.

Principle components analysis is used to show that the dependence of
the expected number of LMXBs per GC on mass, color, and radius is
mainly due to a dependence on a combination of mass and radius, and a
dependence on color. We show that this result implies that the
dependence on mass, color, and size is essentially equivalent to a
dependence on the encounter rate $\Gamma_h$ and the metallicity
$\gdiff{Z}$. The best-fit form is
\begin{equation}
  \lambda_{t} = A \,
                \left(\frac{\Gamma_{h}}
                           {10^{7}}
                \right)^{0.818^{+0.050}_{-0.049}} \,
                \left(\frac{\gdiff{Z}}
                           {\gdiff{Z_{\odot}}}
                \right)^{0.389^{+0.071}_{-0.067}} \,
  .
\end{equation}
This provides strong evidence for dynamical formation playing a
primary role in forming LMXBs in the dense stellar environs of
extragalactic GCs, strengthening the evidence presented in
\citet{JCF+2004}. For our Complete sample, $L_X > 3.2\times 10^{38}
{\rm \, erg \, s}^{-1}$, we find that the normalization is $A =
4.1\times10^{-2}$.

The $\Gamma_{h}$ exponent is consistent with that derived for
low-luminosity X-ray sources in Galactic GCs \citep{PLA+2003}, but
inconsistent with the simple theoretical prediction that the number of
LMXBs per GC be linearly proportional to $\Gamma_h$. The hardening of
binaries could explain the shallower dependence we
observe. Alternatively, our use of $\Gamma_{h}$ as a proxy for the
encounter rate may affect the detailed dependence, particularly if
core-collapsed extragalactic GCs preferentially contain LMXBs. Our
ongoing study of the GCs and LMXBs in Centaurus A will test our use of
$\Gamma_{h}$.

The metallicity exponent is most consistent with either a metallicity
dependent variation in the number of NSs/BHs per GC, such as one might
find in a metallicity dependent IMF \citep{G1987}, or effects from an
irradiation induced wind \citep{MKZ2004}, although more detailed work
developing irradiation-induced wind models is necessary to predict the
metallicity dependence accurately. The magnetic braking model of
\citet{I2006} may be consistent; however, this model needs to be
developed over a range of metallicities.

The combination of ongoing and upcoming deep {\it CXO} observations
and wide-field {\it HST}-ACS observations of NGC 3379, 4278, 4365,
4697, and 5128 will be essential in extending our ability to test the
LMXB-GC connection to lower X-ray luminosities across entire
galaxies. The greater number of GC-LMXBs will more tightly constrain
the GC-LMXB connection and the theoretical models to explain it. The
spatial distribution of GC-LMXBs and field-LMXBs may determine the
level at which GC-LMXBs that have since been removed from their GC
birthplaces contribute to the field population of LMXBs. It would also
be very useful to extend the studies of LMXBs in early-type galaxies
to include more lenticular galaxies. Our understanding of difference
in the GC-LMXB connection between elliptical and lenticular galaxies
is limited by the small number of lenticular galaxies observed with
{\it CXO}. This limits our ability to use LMXBs to trace the
star-formation history of galaxies. A complementary approach is
required that combines deep studies of individual lenticular galaxies,
as in our approved study of NGC 1023, and shallower studies of larger
samples of lenticular galaxies.

\acknowledgments

Support for this work was provided by the National Aeronautics and
Space Administration through {\it Chandra} awards GO3-4099X,
AR4-5008X, GO4-5093X, and GO5-6086X, issued by the Chandra X-ray
Observatory, which is operated by the Smithsonian Astrophysical
Observatory for and on behalf of NASA under contract
NAS8-03060. Support for Program numbers HST-GO-9401,
HST-GO-10003.01-A, HST-GO-10597.03-A, and HST-GO-10582.02-A was
provided by NASA through grants from the Space Telescope Science
Institute, which is operated by AURA under NASA contract NAS5-26555.
This research was also partially supported by the Celerity Foundation
and the F. H. Levinson Fund. G.~R.~S. acknowledges the receipt of
Achievement Reward for College Scientists and Virginia Space Grant
Consortium Aerospace Graduate Research fellowships. P.C. acknowledges
additional support provided by NASA LTSA grant NAG5-11714.

\bibliography{ms}

\end{document}